\documentclass[12pt,titlepage]{article}

\usepackage{afterpage}

\usepackage{listings}

\usepackage{newfloat}
\DeclareFloatingEnvironment[name={Figure}]{mainFigure}
\DeclareFloatingEnvironment[name={Table}]{mainTable}
\DeclareFloatingEnvironment[name={Figure}]{suppFigure}
\DeclareFloatingEnvironment[name={Table}]{suppTable}

\newcommand{\beginsupplement}{%
        \setcounter{suppTable}{0}
        \renewcommand{\thesuppTable}{S\arabic{suppTable}}%
        \setcounter{suppFigure}{0}
        \renewcommand{\thesuppFigure}{S\arabic{suppFigure}}%
     }

\usepackage{graphicx}
\usepackage{endfloat}
\usepackage{amsfonts}
\usepackage{genetics_manu_style}
\usepackage{float}
\restylefloat{figure}
\restylefloat{table}

\usepackage{siunitx}
\usepackage{amsmath}
\usepackage{amssymb}

\usepackage{graphicx}

\usepackage{natbib}
\bibpunct{(}{)}{;}{author-year}{}{,} 

\usepackage[dvipsnames]{xcolor}

\usepackage{setspace}

\usepackage{subfigure}

\usepackage{multirow,bigdelim}

\usepackage[normalem]{ulem}

\newcommand{\St}{\mathcal{S}}
\newcommand{\It}{\mathcal{I}}
\newcommand{\Rt}{\mathcal{R}}
\newcommand{\traj}{\mathcal{V}}

\newcommand{\tree}{\mathcal{T}}

\newcommand{\ajd}[1]{{\color{black} #1}}

\newcommand{\stochCoalSIR}{stochastic coalescent SIR}
\newcommand{\deterCoalSIR}{deterministic coalescent SIR}

\newcommand{\StochCoalSIR}{Stoch. Coal. SIR}
\newcommand{\DeterCoalSIR}{Deter. Coal. SIR}

\newcommand{\stochSIR}{stochastic SIR}
\newcommand{\StochSIR}{Stochastic SIR}
\newcommand{\BDSIR}{BDSIR}

\title{\Large{Inferring epidemiological dynamics with Bayesian coalescent inference:  The merits of deterministic and stochastic models}\\
\vspace{10mm}}

\author{Alex Popinga\thanks{Department of Computer Science, University of Auckland, Auckland, New Zealand 1010}\thanks{Allan Wilson Centre for Molecular Ecology and Evolution, New Zealand PN4442},  
Tim Vaughan$^{\ast}$$^{\dag}$\thanks{Massey University, Palmerston North, New Zealand 4442}, 
Tanja Stadler\thanks{Department of Biosystems Science and Engineering, ETH Z\"urich, Basel, Switzerland}, 
Alexei J Drummond$^{\ast}$$^{\dag}$}

\renewcommand{\CorrespondingAddress}{Department of Computer Science \\
    University of Auckland \\ 
    303 - 379\\
    38 Princes St\\ 
    Auckland, NZ 1010 \\ 
    (64 9) 373-7599 88298 \\
    \texttt{alexei@cs.auckland.ac.nz} \vfill}
\renewcommand{\RunningHead}{Bayesian Coalescent Epidemic Inference}
\renewcommand{\CorrespondingAuthor}{Alexei J Drummond}
\renewcommand{\KeyWords}{Bayesian inference, Phylodynamics, Coalescent, Epidemic, Stochastic} 

\begin{document}

\maketitle

\begin{abstract}

Estimation of epidemiological and population parameters from molecular sequence data has become central to the understanding of infectious disease dynamics.  
Various models have been proposed to infer details of the dynamics that describe epidemic progression. 
These include inference approaches derived from Kingman's coalescent theory. 
Here, we use recently described coalescent theory for epidemic dynamics to develop stochastic and deterministic coalescent SIR tree priors. We implement these in a 
Bayesian phylogenetic inference framework to permit joint estimation of SIR epidemic parameters and the sample genealogy. 
We assess the performance of the two coalescent models and also juxtapose results obtained with BDSIR, a recently published birth-death-sampling model for epidemic inference.  
Comparisons are made by analyzing sets of genealogies
simulated under precisely known epidemiological parameters.  
Additionally, we analyze influenza A (H1N1) sequence data sampled in the Canterbury region of New Zealand and HIV-1 sequence data obtained from
known UK infection clusters. 
We show that both coalescent SIR models are effective at estimating epidemiological parameters from data with large fundamental reproductive number $R_0$ and large population size $S_0$. 
Furthermore, we find that the stochastic variant generally outperforms its deterministic counterpart in terms of error, bias, and highest posterior density coverage, particularly for smaller $R_0$ and $S_0$.  
However, each of these inference models are shown to have undesirable properties in certain circumstances, especially for epidemic outbreaks with $R_0$ close to one or with small effective susceptible populations.

\end{abstract}

\section{Introduction}

\subsection{Phylodynamics and the coalescent}

The epidemiological and evolutionary processes that underpin rapidly evolving 
species occur on a shared spatiotemporal frame of reference.  Unified analyses 
that include both the dynamics of an epidemic and the reconstruction of the pathogen phylogeny can 
therefore uncover otherwise inaccessible information to aid in outbreak prevention.  Such information 
includes the rates of pathogen transmission and host recovery, effective population sizes, and the `time of origin' representing the 
introduction of the first infected individual into a population of susceptible hosts.

The term \textit{phylodynamics} was popularized by \cite{Grenfell16012004} to describe the interlaced study of immunodynamics, 
epidemiology, and evolutionary mechanisms.  Several phylodynamic models, both 
stochastic and deterministic in nature, have since been developed to characterize the 
phylogenetic history of the pathogen species and compartmentalizations of the host 
population throughout the epidemic.  Such models grant the ability to infer 
key epidemiological parameters from genetic sequence data and include birth-death 
branching processes \citep{Stadler:2012,Stadler:2013,Kuhnert:2014,gavryushkina2014bayesian}, as well as 
coalescent approaches \citep{GriffithsandTavare:1994,Pybus22062001, KoelleandRasmussen, Rasmussen2011, DearloveandWilson,Rasmussen2014} derived from 
Kingman's coalescent theory \citep{Kingman:1982}.

Significant steps toward the unification of epidemiology and
statistical phylogenetics were made by \cite{Pybus:2001}, \cite{DearloveandWilson}, and 
\cite{Volz:2009}, with the formalization and application of Kingman's $n$-coalescent to pathogen population
dynamics. These methods involved numerical integration
of a set of ordinary differential equations (ODEs) to find deterministic approximations to the variation in the number of sampled
lineages through time. 
\cite{Volz:2012} extended the tree density calculation from previous work \citep{Volz:2009} to allow for
serially-sampled and spatially structured genetic sequence data.  In
this coalescent model, the birth and death rates can vary in time and
by the state of the host, so that ``the birth rate of a single gene
copy is both time- and state-dependent.''

In this paper, we assess the ability of coalescent-based phylodynamic models to infer, in a Bayesian setting, a range of epidemiological parameters from simulated data.  
While \cite{DearloveandWilson} paved the way by implementing a coalescent approach for deterministic SI, SIS, and SIR models for Bayesian inference, 
we implement and rigorously test both deterministic and stochastic coalescent SIR models of epidemic dynamics extended for heterochronously sampled data.

\subsection{Stochastic and deterministic models}

Stochasticity and determinism in population sizes each maintain dominant roles in particular stages 
of an epidemic.  Once the infected population has grown considerably large, on the order 
of 1,000 to 10,000 lineages, the probability densities of stochastically-expressed 
population size dynamics converge toward the deterministic interpretation \citep{Rouzine:2001}. However, 
during the early stages the population size of infected individuals is small, and the 
dynamics of the epidemic are therefore governed by stochastic processes due to the 
relative significance of fluctuations in the demographic and rate parameters of the 
population model \citep{Kuhnert:2014}.  Therefore, approximating the prevalence of infection by a deterministic function requires the 
number of infected hosts within the effective population to be assumed as very large 
throughout the duration of the described epidemic, i.e., once the exponential growth phase 
has been reached \citep{Rouzine:2001}.

Population size is critical to the epidemiological system and, as with any parameter in a 
Bayesian setting, yields the most accurate estimations when detailed prior information 
is available and incorporated into the inference \citep{Drummond:2006}.   
In our extension and implementation of the coalescent model for epidemics, both stochastic and deterministic population size
processes are used for the simulation of trees and/or trajectories for subsequent inference.

\subsection{Compartmental population models (SIR)}

Host populations can be compartmentalized simply but effectively in 
mathematical models that describe epidemic progression.  The specific division of the 
aggregate population depends on the contagion, spanning a range 
of scenarios where hosts may or may not recover from infection, may or may not be reinfected, etc.  
Such examples include the SI (Susceptible-Infected), 
SIS (Susceptible-Infected-Susceptible), and SIR (Susceptible-Infected-Removed) models \citep{Anderson:1991fk, Keeling:2008zr}.
Each of these compartments can be expressed either
(a) by a set of ODEs that describe the deterministic
time development of real-valued compartment occupancies, or (b) in terms of integer-valued
occupancies governed by continuous-time Markov chains (CTMC) that allow for a degree of
uncertainty in the timing and number of events that occur over the course of the epidemic.


In this paper, we concentrate on the SIR model, which describes epidemics that 
include infected individuals who are at some point in time removed from the effective 
population by way of immunity, death, behavioral changes, or some other termination of infectiousness.  
The deterministic variant of this model was introduced by \cite{KermackandMcKendrick} and is given by the trio of coupled
ODEs,
\begin{eqnarray}
\frac{d}{dt} S(t) &=& - \beta I(t)S(t),\label{eq:SIR1}\\
\frac{d}{dt} I(t) &=& \beta I(t)S(t) - \gamma I(t),\label{eq:SIR2}\\
\frac{d}{dt} R(t) &=& \gamma I(t),\label{eq:SIR3}
\end{eqnarray}
where $\beta$ and $\gamma$ respectively represent the transition rates
from susceptible $S$ to infected $I$, and infected $I$ to removed
$R$. 
\ajd{The model fully defines the population dynamics with initial conditions $S(z_{0})$, $I(z_{0})$, and $R(z_{0})$.  It is worth recognizing that, in the closed SIR model used here, 
there is no demographic change in the host population. Therefore, $\frac{d}{dt} S(t)+\frac{d}{dt} I(t)+\frac{d}{dt} R(t)=0$ 
and $S(t)+I(t)+R(t)=N$, where $N$ is the constant total population size.}
 Throughout this paper we refer to the solutions to eq.~(1-3) as \emph{deterministic SIR trajectories}.

The comparable stochastic description is given in terms of the probability
of the epidemic state at time $t$ given its initial state and the rate parameters
\begin{equation}
\pi(s,i,r;t) \equiv \Pr(S(t)=s, I(t)=i, R(t)=r|S(0),I(0),R(0),\beta,\gamma),
\end{equation}
which is governed by the following equation of motion:
\begin{align}
\frac{d}{dt}\pi(s,i,r;t)=&\beta\left[(s+1)(i-1)\pi(s+1,i-1,r;t)-si\pi(s,i,r;t)\right]\nonumber\\
&+\gamma\left[(i+1)\pi(s,i+1,r-1;t)-i\pi(s,i,r;t)\right].
\label{eq:SIRME}
\end{align}
An explicit sampling process is incorporated by allowing each removal
event to coincide with a sampling event with a fixed probability
$\psi/(\psi+\mu)$ where $\psi$ and $\mu$ are the overall rates of
sampled and unsampled removals, respectively, such that
$\gamma=\psi+\mu$. We refer to epidemic histories sampled from this
model as \emph{stochastic SIR trajectories}.

Both types of epidemic trajectories can be related to models of sampled transmission tree
genealogies. In the deterministic case, this relationship is made via
the coalescent distributions described in 
\cite{Volz:2012}. We call this the
\emph{deterministic coalescent SIR model}.  In the stochastic case,
genealogies appear naturally from a branching process in which the
branching events coincide with the transmission events in the CTMC and
only those lineages ancestral to sampled removals are recorded. We
call this the \emph{stochastic SIR model}.  The \emph{BDSIR model} introduced by
\cite{Kuhnert:2014} provides an approximation to the stochastic
SIR model.

Another way of relating the stochastic SIR model to sampled
transmission trees involves drawing a realization of a stochastic SIR
epidemic, then using the coalescent distribution in \cite{Volz:2012} to
produce a tree conditional on the particular piecewise constant
infected compartment size corresponding to that realization.  We call
this approach the \emph{stochastic coalescent SIR model}.  Unlike BDSIR, the stochastic 
coalescent SIR model does not require the sampling process to be specified explicitly.

Both the transmission rate $\beta$ and removal rate $\gamma$ can be
estimated using each of the methods considered in this paper from data
ascribed to an SIR epidemic.

\section{Methods}

\subsection{Inference framework}

All phylodynamic inference discussed in this paper is based on the joint posterior probability density
\begin{eqnarray}
\ f(\tree,\traj,\eta, \theta | D) = \frac{\Pr (D|\tree,\theta) f(\tree|\traj,\eta) f(\traj |\eta) f(\eta) f(\theta) }{\Pr(D)},
\label{eq:posterior}
\end{eqnarray}
where the sampled transmission tree $\tree$, the epidemic trajectory denoted $\traj = \left( \St,\It,\Rt \right)$, the substitution parameters $\theta$, and the epidemiological parameters $\eta = \{\beta,\gamma,S_0,z_0\}$ are all estimated from the sequence data.  
The sampled transmission tree $\tree$ is assumed to be identical to the pathogen genealogy.

Here, $\St$, $\It$, and $\Rt$ represent the host compartment sizes from the present time $\tau=0$ back to the origin $z_{0}$, such that:  $\St(\tau) = S(z_{0}-\tau)$, $\It(\tau) = I(z_{0}-\tau)$, and $\Rt(\tau) = R(z_{0}-\tau)$. 

The various terms making up the right-hand side of
eq.~(\ref{eq:posterior}) are the tree likelihood
$\Pr(D|\tree,\theta)$, the tree prior $f(\tree|\traj,\eta)$,
the epidemic trajectory density $f(\traj|\eta)$, and the substitution
and epidemiological parameter priors $f(\eta)$ and $f(\theta)$. The
probability $\Pr(D)$ is merely a normalizing constant and can
be ignored.  It is the product of the tree prior and trajectory
density $f(\tree|\traj,\eta)f(\traj|\eta)$ that distinguishes each of
the models considered in this paper.

For both the deterministic and stochastic coalescent SIR models, the tree prior $f(\tree|\traj,\eta)$ is calculated in the following way.  First, consider the time span of a tree divided into segments bracketed by both 
sampling and coalescent events. By considering intervals ending in sampling events as well as coalescent-ending intervals, we follow previous work
that extended coalescent approaches to time-stamped, serially-sampled data \citep{RodrigoFelsenstein1999,Drummond:2002}.
 Interval $i$ is spanned by $k_i$ lineages and is the 
$i$'th interval when ordered from the most recent tip to the root. The set of intervals 
$A$ ending in sample events and the set of intervals $Y$ ending in coalescent events 
together encompass all intervals, $V = A \cup Y$. Let the end time of an interval be $\tau_i$ 
(going back in time), with $\tau_0=0$ as the time of the most recent tip and with time increasing 
into the past. Then the probability density of a genealogy given an epidemic trajectory is
\begin{equation}
f(\tree|\traj, \eta) = \prod_{i \in Y} {\lambda}_{k_{i}}(\tau_i) \prod_{i\in V}\omega(\tau_i, 
k_i),
\label{eq:treeGivenTraj}
\end{equation}
where $\lambda_{k_i}(\tau)$ is the instantaneous coalescent rate at $\tau$ prescribed by
\cite{Volz:2012}
\begin{align}
\lambda_{k_i}(\tau) &= \binom{k_i}{2}\frac{2{\beta}{\St}({\tau})}{{\It}({\tau})},
\label{eq:coalrate}
\end{align}
and where $\omega(\tau_i,k_i)$ is the survival probability
\begin{equation}
\omega(\tau_i, k_i) = \exp\left({-{\int_{\tau_{i-1}}^{\tau_{i}}{\lambda_{k_i}(\tau)d\tau}}}\right).
\end{equation}

The deterministic coalescent SIR model assumes that the SIR epidemic
trajectories are found by integrating the ODEs in
eqs.~(\ref{eq:SIR1})--(\ref{eq:SIR3}).  Therefore, under this model each
epidemic trajectory is a deterministic function of its parameters
$\traj(\eta)$.  This means that the trajectory density can be written as
\begin{equation}
f(\traj|\eta)=\delta(\traj-\traj(\eta)),
\end{equation}
where $\delta(x)$ is the Dirac delta function and represents a point
mass concentrated at $x=0$.

In contrast, the stochastic coalescent SIR model assumes that the
epidemic is generated by a jump process corresponding to the master
equation given in eq.~(\ref{eq:SIRME}). In this case, the probability
$f(\traj|\eta)$ is nonsingular and thus contributes to the uncertainty
in the final inference result.

In the BDSIR model, $f(\traj|\eta)$ is the same as for the stochastic
coalescent SIR model, but $f(\tree|\traj,\eta)$ is defined
differently.  See \cite{Kuhnert:2014} for details.

\subsection{MCMC algorithm}

We use Markov chain Monte Carlo (MCMC) to sample from the joint
posterior density given in eq.~(\ref{eq:posterior}).  Many of the
specifics of the algorithm used have been discussed previously; in
particular the method for calculating the tree likelihood
\citep{Felsenstein:1981,Felsenstein:2004} and mechanism for exploring
tree space \citep{Drummond:2002}. However, the model-specific product
$f(\tree|\traj,\eta)f(\traj|\eta)$ requires special attention.

As we are primarily interested in parametric inference rather than the
epidemic trajectory itself, we can regard $\traj$ as a nuisance
parameter to be marginalized over. This marginalization can be
achieved implicitly by sampling it using MCMC and then ignoring this component of the sampled state, 
which is the strategy we use when reporting the BDSIR results. It can
also be made an explicit part of the likelihood calculation, which is the approach
we take with the deterministic and stochastic coalescent SIR
models. This marginalization means that the product
$f(\tree|\traj,\eta)f(\traj|\eta)$ becomes
\begin{equation}
f(\tree|\eta)=\int f(\tree|\traj,\eta)f(\traj|\eta)\mathrm{d}\traj,
\label{eq:treeprior}
\end{equation}
the probability density of the tree given the epidemiological parameters.

In the case of the deterministic coalescent SIR model, this density
reduces to $f(\tree|\traj(\eta),\eta)$, meaning that the density of
the tree given epidemiological parameters $\eta$ is obtained simply by
substituting the numerical solution to
eqs.~(\ref{eq:SIR1})--(\ref{eq:SIR3}) for those parameters into
eq.~(\ref{eq:treeGivenTraj}).

The stochastic coalescent SIR model is more complex, as in this case
the trajectory density $f(\traj|\eta)$ is nonsingular, meaning that
computing the integral in eq.~(\ref{eq:treeprior}) is nontrivial. We
treat this here using the ``pseudo-marginal''
approach \citep{Beaumont:2003,Andrieu:2009} in which, at each step in
the MCMC chain, the marginalized tree density $f(\tree|\eta)$ is
replaced by the Monte Carlo estimate
\begin{equation}
\hat{f}(\tree|\eta)=\frac{1}{M}\sum_{r=1}^{M}f(\tree|\traj_r,\eta),
\end{equation}  
where each $\traj_r$ is a trajectory sampled independently from
$f(\traj|\eta)$ using a stochastic simulation algorithm
\citep{Sehl:2009aa}.  Perhaps counterintuitively within an MCMC framework, this stochastic likelihood
converges to the true marginal posterior distribution regardless of
the number $M$ of realizations used in the estimate.  However, the
magnitude of $M$ can significantly affect the rate at which the chain
produces effectively independent samples from the posterior and 
must be tuned carefully.

\subsection{Implementation and validation}

We have implemented the schemes described above for performing
inference under the deterministic and stochastic coalescent SIR
models within the BEAST 2 phylodynamics package found at \texttt{http://github.com/CompEvol/phylodynamics}. 
This has a number of advantages over a
stand-alone implementation. Foremost, we were able to
avoid reimplementing components of the algorithm that are in common
with other already-implemented phylogenetic and phylodynamic analyses,
such as the MCMC proposal operators used to traverse the parameter
space.  Furthermore, this greatly increases the usefulness of the
implementation, as it can be immediately used in conjunction with a
wide variety of nucleotide and amino acid substitution models and
parameter priors.

We have taken two steps in order to ensure our implementation
is correct. First, we have compared tree probability density
$f(\tree|\traj, \eta)$ values calculated using the main implementation
of each of the two models with those calculated using completely
independent implementations in R \citep{R}.

Second, we have used the implemented MCMC algorithms to sample
transmission trees from the tree density given in
eq.~(\ref{eq:treeprior}) for each model. We then compared the
distributions of tree height, total edge length, and binary clade count
summary statistics from these sampled ensembles with sample
distributions obtained directly via
stochastic simulation.  As shown in Section 1 (\textit{Sampling from the prior}) 
in the online supporting information, and in the associated figures, the resulting pairs of
distributions agree, providing strong support for our claim that the
implementations of the methods described above are correct.

Instructions for downloading and using this package are also available on
the project web site located at \texttt{http://github.com/CompEvol/phylodynamics}.

\subsection{Simulation study}

To evaluate the implementation and extension of the coalescent models, we performed analyses on both sequence data and fixed trees simulated with 
known parameter values.  The median estimated values produced by each model 
were then used to measure relative error and bias, along with the widths and coverage of 95\% highest posterior density (HPD) intervals.

We used three methods for simulating the trees and trajectories, as shown below:  
\vspace{3mm}
\begin{center}
\begin{tabular}{llll}
\it{Inference model} & \rdelim\{{1}{1mm}[\hspace{1mm} \StochCoalSIR{}] & \DeterCoalSIR{} & \hspace{7.5mm} \BDSIR{} \\
& \hspace{3mm} -------------------------- & ------------------------ & ------------------------ \\
& \rdelim\{{3}{3mm}\hspace{2mm} \StochCoalSIR{} & \StochCoalSIR{} & \StochCoalSIR{} \\
\it{Simulation scheme} & \hspace{5mm} \DeterCoalSIR{} & \DeterCoalSIR{} & \DeterCoalSIR{} \\
& \hspace{5mm} {\StochSIR{}} &  {\StochSIR{}} & {\StochSIR{}}\\
\end{tabular}
\end{center}
\vspace{3mm}
The stochastic coalescent and deterministic coalescent simulation schemes were 
used to validate the coalescent SIR inference models.  The \textit{stochastic SIR} scheme, contrarily, is emphasized for its realistic properties.

Stochastic SIR trees and trajectories were generated using master equations in the simulation package MASTER \citep{Vaughan:MASTER}.  Deterministic coalescent trajectories were generated using a Runge-Kutta integrator \citep{Runge:1895,Kutta:1901} with adaptive step sizes to solve a 
system of first order ODEs.  Stochastic coalescent trajectories were generated using 
Sehl \textit{et al.}'s (2009) SAL tau-leaping algorithm \citep{Sehl:2009aa}.   

To simulate the \stochCoalSIR{} trees, we used the \textit{stochastic SIR} 
trajectories, which could be converted to effective population size 
with the mathematical expression used to obtain Volz's (2012) coalescent rate for the SIR 
model:  $N_e(\tau) = 1/ \lambda_2(\tau) = \It(\tau)/(2{\beta}\St(\tau))$. 
  The sampling times, generated by a sampling rate $\psi$, for the \stochCoalSIR{} trees were also 
taken from the MASTER output to allow for direct comparison between the sets of trees.  
In other words, the underlying epidemic function was the same for both \stochSIR{} and 
\stochCoalSIR{} trees, the latter of which were then simulated under a piecewise constant population function.

Likewise, for the simulation of deterministic coalescent trees we used deterministic SIR 
trajectories to construct a population function and the relation $N_e = \It/(2{\beta}\St)$ to convert 
infected and susceptible host population sizes to effective population size.  The sampling times were randomly generated 
from a probability distribution so that the density of samples taken through time were 
proportional to the number of infected individuals through time, as with the \stochSIR{} trees.

We simulated stochastic SIR trees using multiple combinations of parameter values.  We were particularly interested in varying the basic reproductive ratio $R_0$ and the initial susceptible population size $S_0$, to observe the changes in relative error, bias, and uncertainty in stochastic and deterministic models. 
To alter the ratio ${R_0} = \frac{\beta S_{0}}{\gamma}$ and still generate sensible trees with a consistent number of tips, one or more of the other parameters (birth rate $\beta$, removal rate $\gamma$, or $S_0$) must also change.  
Table 2, as well as Tables S6, S7, and S9 in the supporting information, show the true values of the parameters for each set of simulations.  (The birth rate $\beta$ is not shown, as our implementation allows either $\beta$ or $R_0$ to serve as a parameter in the inference, and $R_0$ is the parameter of interest.  However, $\beta$ can 
be calculated via the other three, using $\beta = \frac{R_{0} \gamma}{S_{0}}$.  For example, when ${R_0} = 1.0978$, ${S_0}=499$, and $\gamma=0.25$, then $\beta=5.50$\mbox{\sc{e}-4}.)

\subsubsection{Heterochronous trees}  We generated 100 trees under each of the three (\stochSIR{}, \stochCoalSIR{}, \deterCoalSIR{}) models
with parameters $S_{0}$, $\beta$, and $\gamma$.  For heterochronously sampled trees, each removal generates a sample with probability $\psi/(\psi+\mu)$, where $\psi$ is the overall rate of sampled removals and $\mu$ is the rate of unsampled removals such that $\gamma = \psi + \mu$.  

The simulations ended once the number of infected individuals reached zero, i.e., when the last infected individual was removed. This ensured that the simulated trajectories spanned past the exponential growth phase of the epidemic and therefore included samples past the peak of infected individuals.  This choice of procedure was motivated by (a) the suggestion of \cite{Stadler:2014} that the behavior of the 
coalescent beyond the exponential phase could either inflate or reduce bias, and (b) the observations of \cite{DearloveandWilson} and \cite{Veronika} that deterministic coalescent SIR models might be properly fitted only once the epidemic has peaked.  
Figure \ref{fig:SIRRh_traj} shows trajectories of susceptible, infected, and removed individuals underlying the simulation of stochastic SIR trees (Figure \ref{fig:Tree}) generated in MASTER.  An example XML for simulating these MASTER trees is provided in the supporting information.

We required that the trees had $n\geq 100$ leaves, filtering out those in which the epidemic died out in the early stages, i.e., when the initial infected individual was removed from the effective population too quickly to infect others.
(Note that the inference procedures
discussed in this manuscript all implicitly condition on the number of leaves.)
The probability that the first event in a given trajectory is the removal (by recovery, death, etc.) of patient zero is given by $\delta/(\beta S_{0} + \delta) = {1}/({1+{R_{0}}})$.
When $R_{0}\approx 2.50$, this probability is $\approx 30\%$.
In our case, 52/152 ($\approx 34\%$) trees were ``empty", or containing only one node.
The filtering process left us with a mean of $\approx 160$ leaves for the simulated trees.

\subsubsection{Homochronous trees}  A major concern in the comparison between \cite{Kuhnert:2014}'s birth-death-sampling SIR inference model, which includes explicit sampling, and our implementations of \cite{Volz:2012}'s coalescent SIR models, which do not include explicit sampling, 
is that the former is given extra information via the sampling process.  
\cite{VolzFrost:2014} have addressed this issue by providing a coalescent SIR model that does incorporate sampling explicitly.

That being said, results from \cite{Veronika} indicate that the poor performance of the deterministic coalescent SIR model in comparison with birth-death models was due to the lack of handling 
stochastic population size changes through time rather than the lack of information about the sampling proportion.  
Their results showed that the coalescent is ``very robust to changes in sampling schemes".

Regardless, to ensure a fair comparison of BDSIR and the coalescent SIR models, we simulated an SIR epidemic with homochronous, or contemporaneous, sampling.  This type of simulation affords no additional information about the population size for explicit-sampling models, as there is only a single time of sampling. 

We selected a simulation time of $t=20$ for the homochronously sampled trees, with the trajectories being sampled at high prevalence but also past the time of peak prevalence.  This is important for distinguishing SIR from SI/SIS outbreaks, as it provides information about the removal parameter $\gamma$.
In this set of simulations, each lineage was sampled at $t=20$ with probability 0.7, (the leaf count distribution for varied sampling probabilities is in the supporting information).

\subsubsection{Simulated sequences}
To assess the ability of each SIR model to infer epidemic parameters with the inclusion of phylogenetic uncertainty, we also simulated the evolution of 2000\,bp sequences down each simulated tree.  
We time-stamped the sequences with the tip dates of each corresponding tree and informed the inference with the true Hasegawa-Kishino-Yano (HKY) substitution model \citep{Hasegawa:1985}, clock rate $=5\mbox{\sc{e}-3}$, and $\kappa=5$.  
These choices were made to reflect real data, specifically that of influenza \citep{Vaughan:2014}. 

Along with simulated sequence data, analyses were performed with the simulated trees fixed (results are in the supporting information), and the parameters $R_0$, $\gamma$, $S_{0}$, and the origin of the tree $z_0$ were 
estimated with Bayesian prior distributions as listed in Table~\ref{table:priors}.

\subsubsection{Deterministic coalescent SIR on higher $R_0$ and $S_0$}  
Finally, we had particular interest in the effects of varying the population size parameter $S_0$ on the deterministic coalescent SIR model, as comparisons from initial analyses with lower true $R_0$ ($\approx$1.5 and $\approx$1.1) and $S_0$ ($=$499) showed higher error and bias and lower 95\% HPD coverage. 
Also, it is often assumed that deterministic descriptions will perform well for higher $R_0$ and larger population sizes.  
Tables 7 and 9 in the supporting information detail the parameter values we used to explore the behavior of the deterministic coalescent on varied $R_0$ and $S_0$ combinations.

\subsection{Interpretation of results}

We compared the coalescent SIR, as well as \BDSIR{}, parameter estimations from the simulated data to the true values 
used to generate the SIR trajectories.
Following \cite{Kuhnert:2014}, the precision and accuracy of these methods were measured by 
relative error, bias, and highest posterior density (HPD) intervals.  We used the posterior 
median value of the parameter value $\overset{\wedge}{\eta}$ compared with the true parameter 
$\bar{\eta} \in\{R_{0}, \gamma, S_{0}, z_{0}\}$.  Relative error and bias are then gauged by calculating the 
median value over medians from all 100 trees, such that
$$RE_{\overset{\wedge}{\eta}} = \frac{\sum\nolimits_{\tau = 1}^{100}\frac{|\overset{\wedge}{\eta} - 
\bar{\eta}|}{\bar{\eta}}}{100} $$

\begin{center}
and
\end{center}
$$RB_{\overset{\wedge}{\eta}} = \frac{\sum\nolimits_{\tau = 1}^{100}\frac{\overset{\wedge}{\eta} - 
\bar{\eta}}{\bar{\eta}}}{100} . $$

\vspace{3 mm}
\noindent{Measures of HPD interval widths are given by}
\vspace{1 mm}
$$\frac{95\% \text{ HPD upper bound} - 95\% \text{ HPD lower bound}}{\bar{\eta}} . $$

Tables 1-3 show these results, along with the percentages of posterior estimates that produced 95\% HPD intervals containing the true values (i.e., 95\% HPD coverage).

\subsection{H1N1 data analysis}

To test the efficacy of the coalescent SIR models on real data, epidemic parameters $R_0$, $\gamma$, $S_0$, and time of origin $z_{0}$ were estimated from 42 seasonal influenza A (H1N1) sequences sampled throughout the 2001 flu season in Canterbury, New Zealand.

Influenza infections are well known for their seasonal SIR behavior in non-equatorial populations, as each annual flu season begins with a supply of susceptible hosts and tapers off as the hosts recover with adaptive immunity \citep{Iwasaki}.  
Due partly to this seasonal pattern, the influenza virus is both a motivator for the development of specialized models as well as a prime subject for testing phylodynamic models \citep{Koelle:2006}.

Sampling a particular region bypasses the necessity of specifying geographically-structured populations, and New Zealand 
is an area of particular interest due to its geographic location and relative isolation from other regions with potentially varying dynamics.  
It is also assumed to play a key role in the global circulation of influenza strains \citep{RambautAndHolmes, Bedford:2010}. 

We used an HKY nucleotide substitution model, with a substitution rate of 5\mbox{\sc{e}-3} as estimated in \cite{Vaughan:2014}, and informed the models with dated sequences.  Priors used for the Bayesian inference are shown in Table \ref{table:priors}.

\subsection{HIV-1 data analysis}
  
In addition to our analysis of H1N1 sequence data, we selected HIV-1 subtype B nucleotide sequences collected from infected individuals located 
in the UK.  The coalescent SIR results were collated with the results from the \BDSIR{} data analysis performed by \cite{Kuhnert:2014} using the same sequences.  More 
details of this analysis are provided in the supporting information.

\section*{Results and Discussion}

\subsection{Simulation study}

Results for epidemic parameter inference from nucleotide sequences simulated from stochastic SIR trees are provided in Table 1 for $R_{0}\approx2.50$. 
Results for inference from fixed trees ($R_{0}\approx2.50$, $R_{0}\approx1.50$, $R_{0}\approx1.10$) are shown in Table 2.
Inference results for analyses with true $R_{0}=1.0987$ and varying population size ($S_{0}=499, 999, 1999$) are described in the supporting information, along with results from trees simulated under the stochastic and deterministic coalescent models for validation.

\subsubsection{Heterochronous trees} 
For $R_0\approx2.50$, all three inference methods performed similarly for parameters $R_0$ and $\gamma$, with high 95\% HPD coverage and low error and bias.  
The most weakly identifiable parameter $S_0$ yielded the largest HPD intervals for all three inference models.  The deterministic coalescent returned higher error (0.52) and bias (0.29) than \stochCoalSIR{} (0.19, -0.03) and \BDSIR{} (0.39, 0.24) 
and recovered the origin parameter $z_0$ for only 76 out of 100 simulated trees, while the stochastic coalescent and \BDSIR{} respectively recovered $z_0$ for 99 and 97 out of 100 simulations.    

For $R_{0}\approx1.50$, the relative HPD widths (akin to variance) for three of the four estimated parameters ($R_0$, $\gamma$, and $z_0$) were smallest for \BDSIR{}.  For the 
parameter $S_0$, the relative HPD width is largest for \BDSIR{}, although it also had slightly higher 95\% HPD coverage than \deterCoalSIR{} and the same as \stochCoalSIR{}.  
The \deterCoalSIR{} method recovered the truth for 85, 89, 91, and 88 out of 100 trees for parameters $R_0$, $\gamma$, $S_0$, and $z_0$, while its stochastic analog recovered the truth for 100, 85, 100, and 99 out of 100 trees for the same parameters.
Finally, for \stochCoalSIR{} and \BDSIR{}, error and (absolute) bias were relatively low for $R_{0}$, arguably the parameter of most interest to epidemiologists since it represents the number of individuals each infected individual will themselves infect in a naive population.  
Deterministic coalescent SIR has a higher error (0.24) and bias (0.15) and also has significantly lower coverage for $R_0$ (85\%).   

For $R_{0}\approx1.10$, the two stochastic models again outperformed the deterministic coalescent in error, bias, and 95\% HPD coverage.  
The stochastic coalescent most reliably recovered the truth for $R_0$ (99 out of 100 simulations), while the deterministic coalescent had more than double the error and bias and 
still only recovered the truth for 25 of the 100 simulations.  
\BDSIR{} had the lowest error and bias for $R_0$ under this scheme, although it only recovered the truth for 75 out of 100 simulations.
For removal parameter $\gamma$, \BDSIR{} again yielded lower error and bias, in this case returning the truth for 100/100 trees (in contrast to 84 and 86 from the stochastic and deterministic coalescent, respectively).

In the stochastic models, there 
is a greater tradeoff between parameters due to the impact the relationship between them has on the survival of trajectories at low $R_0$.  
A larger estimated removal rate tends to require a larger susceptible population in order for the epidemic to avoid dying out in the early 
stages.  Likewise, a smaller susceptible population implies a smaller estimated $\gamma$.

\subsubsection{Deterministic coalescent SIR on higher $R_0$ and $S_0$}  As mentioned in the preceding subsection, 
the deterministic coalescent model yielded higher error and bias than both the stochastic coalescent and \BDSIR{} for most parameters with $R_{0} \approx 1.10$ and $S_{0}=499$.  

To investigate the deterministic model's sensitivity to population sizes, we also simulated a range of population sizes ($S_{0}=$ 499, 999, and 1999) for $R_{0}=1.0987$.  
Even with $S_{0}=1999$, the \deterCoalSIR{} model's 95\% HPD coverage was low.  For parameters $R_0$, $\gamma$, $S_0$, and $z_0$, 
this coverage was respectively: 40\%, 64\%, 66\%, and 18\%.  Table S6 in the supporting information shows these results.  

Additionally, we increased both $R_0$ (to 3.5 and 5) and $S_0$ (to 4999 and 9999).  However, for parameters $R_0$, $\gamma$, and $S_0$, the \deterCoalSIR{} showed increased error, bias, and HPD widths, 
and the HPD coverage for $z_0$ did not improve.  These results are shown in Table S9 in the supporting information.

While each of these methods are approximations, the deterministic coalescent particularly suffers from model misspecification since it does not account for the stochasticity that is always present in the early stages of epidemics, regardless of $S_0$.

\subsubsection{Homochronous trees} Results for homochronously sampled trees are given in Table S3 in the supporting information.  

All three SIR inference models recover the truth for more than 95/100 trees within their respective 95\% HPD widths for epidemic parameters $R_{0}$, $\gamma$, and $S_{0}$.  The time of origin $z_{0}$ was recovered for 100/100 trees by BDSIR, 95/100 
trees by stochastic coalescent SIR, and 73/100 trees by deterministic coalescent SIR.  However, relative error and bias also increased consistently across all three models, along with the 95\% HPD widths.  
The deterministic coalescent had the highest error, bias, and HPD width for $R_0$ and highest error and HPD width for $S_0$, which is consistent with the heterochronously sampled data.  

Further consideration of the effects of sampling rate changes and sampling model misspecification are warranted for BDSIR and coalescent SIR, the latter of which has been facilitated by \cite{VolzFrost:2014}.

\subsubsection{Simulated sequences} Relative error and bias were inflated across all three inference models with the addition of phylogenetic uncertainty, and in certain cases the 95\% HPD coverage was lower than with fixed trees.  The 
deterministic coalescent model only recovered the truth within its 95\% HPD intervals for 90 or more of the 100 trees in the case of $S_0$.  The true values for the parameters $R_{0}$, $\gamma$, and $z_{0}$ were covered by 95\% HPD intervals for 87, 56, and 29 of the 100 trees, respectively.  
This is contrasted with the performance of the stochastic coalescent (100, 97, 47, and 37 for parameters $S_{0}$, $R_{0}$, $\gamma$, and $z_{0}$) and BDSIR (99, 100, 84, and 18 for $S_{0}$, $R_{0}$, $\gamma$, and $z_{0}$), as shown in Table 1.

Error, bias, and 95\% HPD widths were higher with simulated sequences for all three inference models for parameters $\gamma$, $S_0$, and $z_0$, than with fixed trees.  
This indicates the importance of calibrating epidemic parameters of interest.  In our case, we emphasize the basic reproductive number $R_0$, often the parameter of most interest to epidemiologists.  
For $R_0$, \stochCoalSIR{} and \BDSIR{} recovered the truth within their 95\% HPD intervals for 97 and 100 of the 100 simulations, respectively.  
They also showed only slight changes in error and bias compared to inference performed on the fixed trees used to generate the sequences.  
The \deterCoalSIR{} model recovered $R_0$ for 87 of the 100 simulations (contrasted with 98/100 for the fixed trees), and with increased error.

\subsubsection{Priors and identifiability} It is important to understand the impact of selected priors on inference results, as the priors are where the power of Bayesian inference lies.  
For example, we found relatively weak identifiability in the initial susceptible population parameter $S_0$, which must either be fixed or estimated alongside the origin parameter $z_0$.  

In addition to allowing each parameter to be either fixed or estimated, we have provided options for parameterization of our models, with either the transmission rate $\beta$ or $R_0$ acting as operable parameters in MCMC analysis.  For the deterministic coalescent, there is also an option to use the 
intrinsic growth parameter described by \cite{DearloveandWilson}.  

The choice of parameterization necessarily affects the prior that will be used in the inference and should be considered carefully.  However, we found that once a parameterization has been selected, our inference models are robust to different prior distributions placed on each parameter.  
We also used broader prior distributions on the deterministic coalescent to test whether this would increase its lower 95\% HPD coverage relative to the stochastic models.  We found that doing so increased the error and bias of the results without increasing the accuracy, (shown in Table S4 in the supporting information).

\subsection{H1N1 data analysis}

Epidemic parameter estimates from serially-sampled influenza A (H1N1) virus
sequence data are shown in Table \ref{table:H1N1}.  

The estimated means of the basic reproductive number were $R_{0}=$1.46, 1.35, and 1.61
for the stochastic coalescent, deterministic coalescent, and BDSIR, respectively.  
Estimates of $R_0$ from pandemic H1N1 in New Zealand range from about 1.2 to 1.5 
\citep{Paine:2010,RobertsandNishiura:2011,Opatowski:2011,Roberts:2013,Biggerstaff:2014}, and 
estimates of $R_0$ for seasonal H1N1 from other countries also range from around 1.2 to 1.5 \citep{Chowell:2008}.
The 95\% HPD intervals were very similar across each model, ranging from just over 1.0 to around 2.0.  

The population of the Canterbury region in 2001 was reported to be around $481,431$ by the
Environment Canterbury Regional Council \citep{ECAN} and $521,832$ by
Statistics New Zealand \citep{StatsNZ}. The mean estimates of $S_0$ were considerably lower using the stochastic coalescent ($S_{0}=69,000$), 
the deterministic coalescent ($S_{0}=120,000$), and BDSIR ($S_{0}=22,200$).  However, the \textit{effective} population of susceptibles 
is assumed to be much smaller, as the total population contains individuals of various susceptibility, e.g., those with partial immunity from vaccination and previous or secondary infections.  

Most people recover from flu symptoms, the time they are likely to be most infectious, within a few days up to two weeks \citep{CDC,WHO}.  
This provides a range of probable true values for the removal parameter $\gamma$.  The sequence data and molecular clock rate, and therefore 
the tree, are in units of years.  Therefore, our $\gamma$ range would be 365/14 days to 365/2 days, or $\gamma=26.1$ 
to $\gamma=182.5$.  The stochastic coalescent, deterministic coalescent, and \BDSIR{} respectively inferred $\gamma$ means of:  27.08, 34.50, and 27.72.  
These estimates are on the low side compared to epidemiological models for influenza that include explicit spatial and household effects \citep{Ferguson:2005}, 
but a moderate misfit of the model is not unexpected when fitting a simple closed SIR model with no population substructure.

The root of the tree was very similar across all inference models,
respectively:  0.53, 0.54, and 0.49 for stochastic coalescent SIR,
deterministic coalescent SIR, and BDSIR.  The same was true for the origin $z_0$, with:  0.69, 0.73, and 0.53 for the stochastic coalescent, deterministic coalescent, and BDSIR.
All three inference models returned tree root and origin estimates that
are consistent with previous estimates from single flu seasons.  That
is, the tree age is young and the root coincides with the start of the (winter) influenza season in the Southern Hemisphere.  
The time of introduction of influenza into the region, $z_0$, was one or two months before the root.  This supports the notion that the sequences selected represent a single introduction of the strain into the Canterbury population (see supporting information for details of data selection).

The trees estimated by each of the three models are
typical for influenza (see Figure 11 for representative trees from each posterior), with branches that are quick to coalesce moving backward
in time from the most recently sampled tip.

\subsection{HIV-1 data analysis}  

Results for inference from HIV-1 sequence data can be found in the supporting information.

\subsection{Computational efficiency}  Finally, we supply Table S5 in the supporting information to show comparisons of computation times under each inference model for each type of data analyzed.  
The \deterCoalSIR{} model is by far the fastest to sample and converge, with \stochCoalSIR{} and \BDSIR{} varying depending on the type of data.

\subsection{Closing remarks}

A key reason for the success of coalescent theory in population genetics is its mathematical simplicity and
the computational efficiency of calculating the probability density of a sample genealogy.
Our results show that a stochastic variant of coalescent theory can be successfully adapted to estimate epidemiological parameters in a true Bayesian inference context.
This stochastic coalescent SIR model performs better than the deterministic analog for estimating epidemic parameters in some circumstances.
Unfortunately, the stochastic model relies on a computationally demanding Monte Carlo estimate of the coalescent density via simulation of an ensemble of epidemic trajectories, negating one of the main advantages of coalescent theory.
In fact, the current implementation is less computationally efficient than the implementation of the BDSIR model. 
However, an advantage of the stochastic coalescent over the explicit sampling model in \BDSIR{} is its robustness to biased sampling schemes, as has been shown for the case of pure exponential growth dynamics \citep{Veronika}.

A more computationally efficient approach to computing the coalescent probability of the sample genealogy in the stochastic setting would be to use particle filtering \citep{Andrieu:2009,andrieu2010particle,Rasmussen2011,Rasmussen2014}, but there are no theoretical barriers to applying particle MCMC to the exact model \citep{Stadler:2014}.
Therefore, an obvious extension of this work would be to apply particle MCMC algorithms to the exact stochastic SIR model that was used in simulations in this current work. We would anticipate that the exact model would outperform all the methods tested here, especially when $R_0$ is close to one.

In the meantime, the Bayesian coalescent inference methods developed here make it feasible to estimate epidemic parameters from time-stamped, serially-sampled molecular sequence data, while accurately accounting for uncertainty in the topology and the divergence times of the phylogenetic tree.

\section{Acknowledgments}
AJD was funded by a Rutherford Discovery Fellowship from the Royal Society of New Zealand. 
AP, TV, TS and AJD were also partially supported by Marsden grant \#UOA1324 from the Royal Society of New Zealand. (http://www.royalsociety.org.nz/programmes/funds/
marsden/awards/2013-awards/) 

\begin{mainFigure}[H]
    \centering
    	\includegraphics[width=6in]{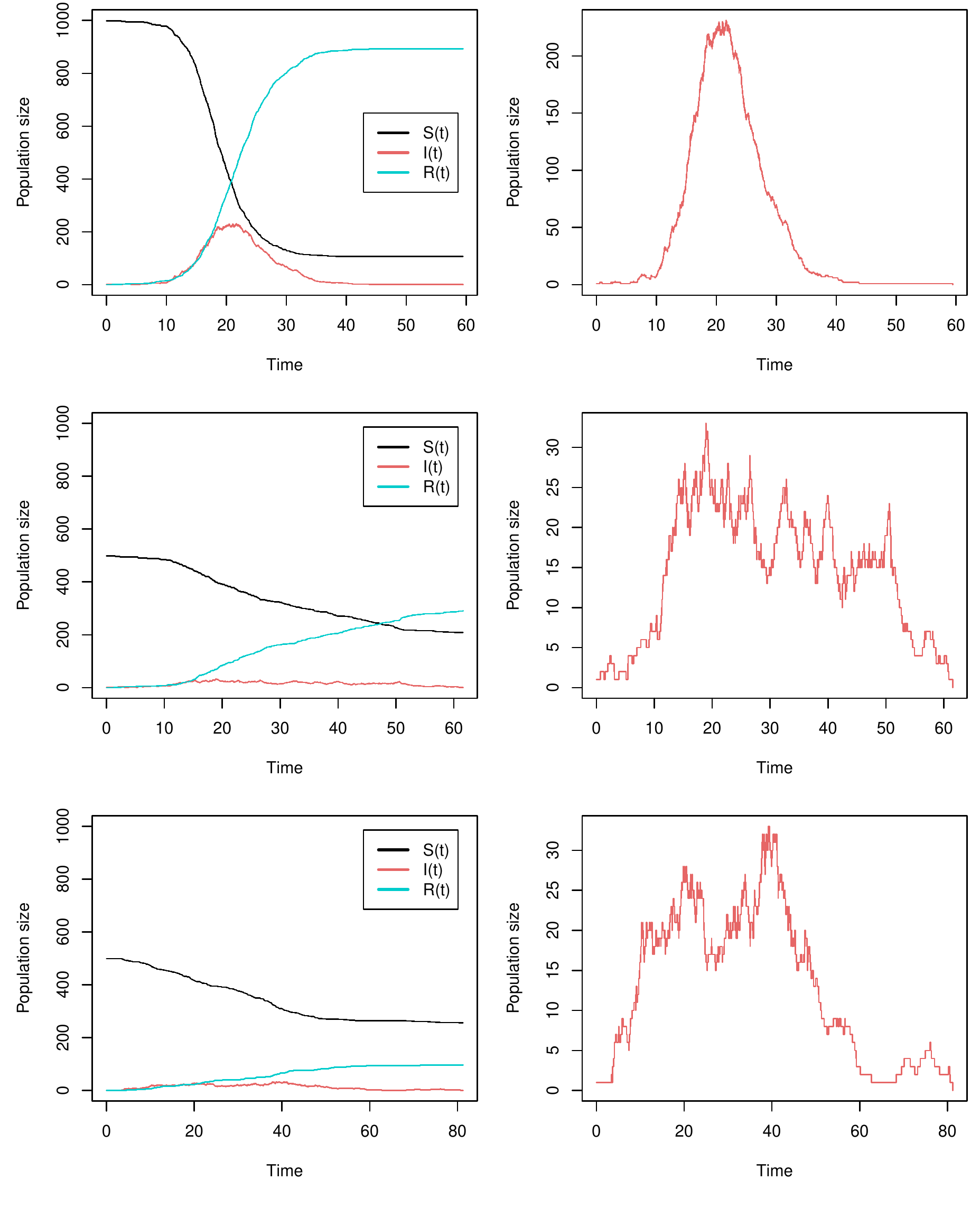}
    \caption{Stochastic SIR trajectories for susceptible $S$, infected $I$, and recovered $R$ populations, with (top row) 
    $S_0=999$ and $R_0=2.4975$, (second row) $S_0=499$ and $R_0=1.497$, and (bottom row) $S_0=499$ and $R_0=1.0978$.  (The 
    second column shows infected $I$ only.)}
	\label{fig:SIRRh_traj}
\end{mainFigure}
\begin{mainFigure}[H]
  \begin{center}
    \includegraphics[width=4in]{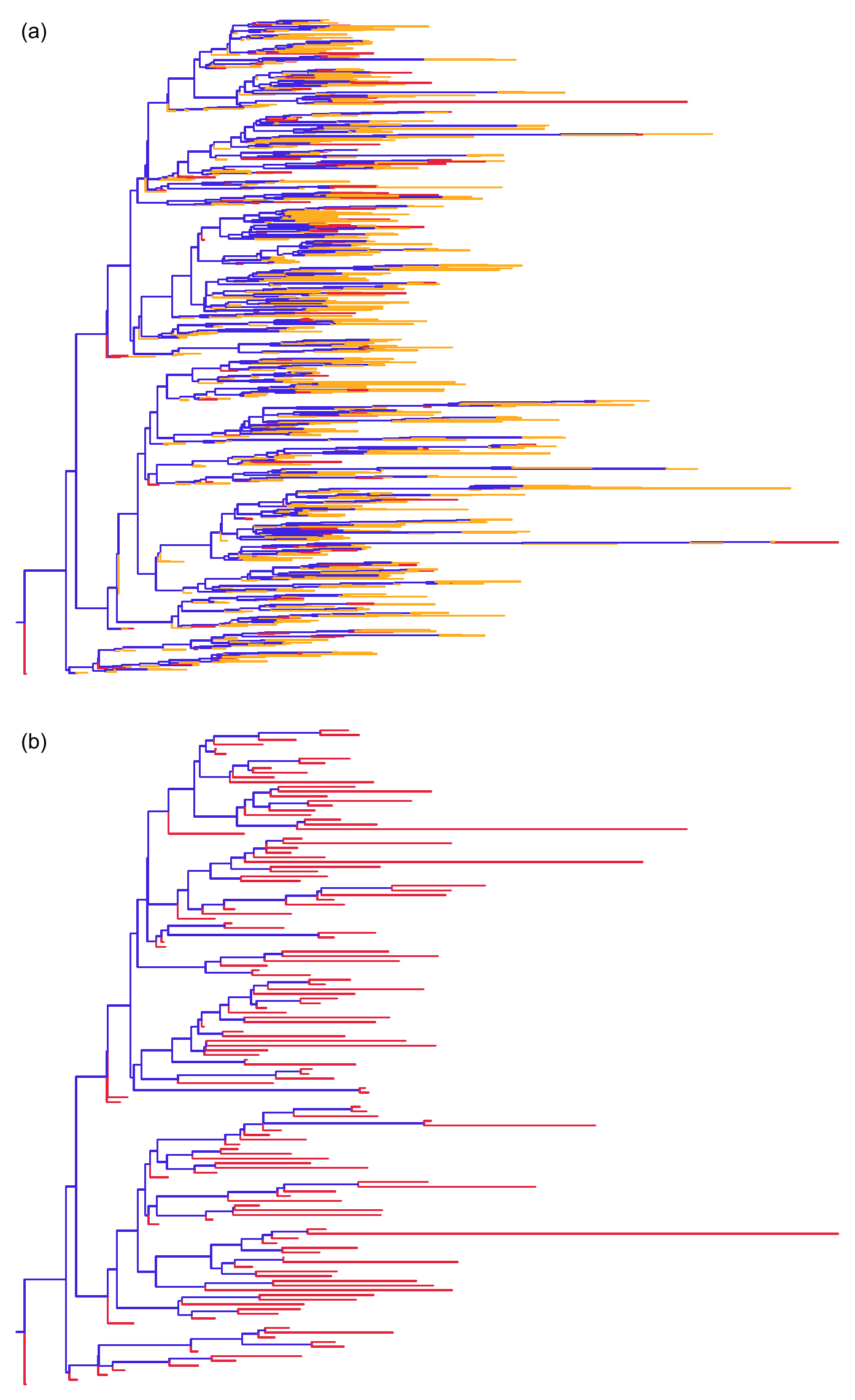}
  \end{center}
    \caption{(a) Full stochastic SIR \textit{transmission} tree with both sampled $\psi$ tips, shown in red, and otherwise removed $\mu$ tips, shown in yellow. 
      (b) The corresponding 140-tip \textit{sampled} stochastic SIR tree. 
      Figures generated in FigTree \citep{FigTree}.}
  \label{fig:Tree}
\end{mainFigure}
\begin{mainFigure}[H]
    \centering
    	\includegraphics[width=7in]{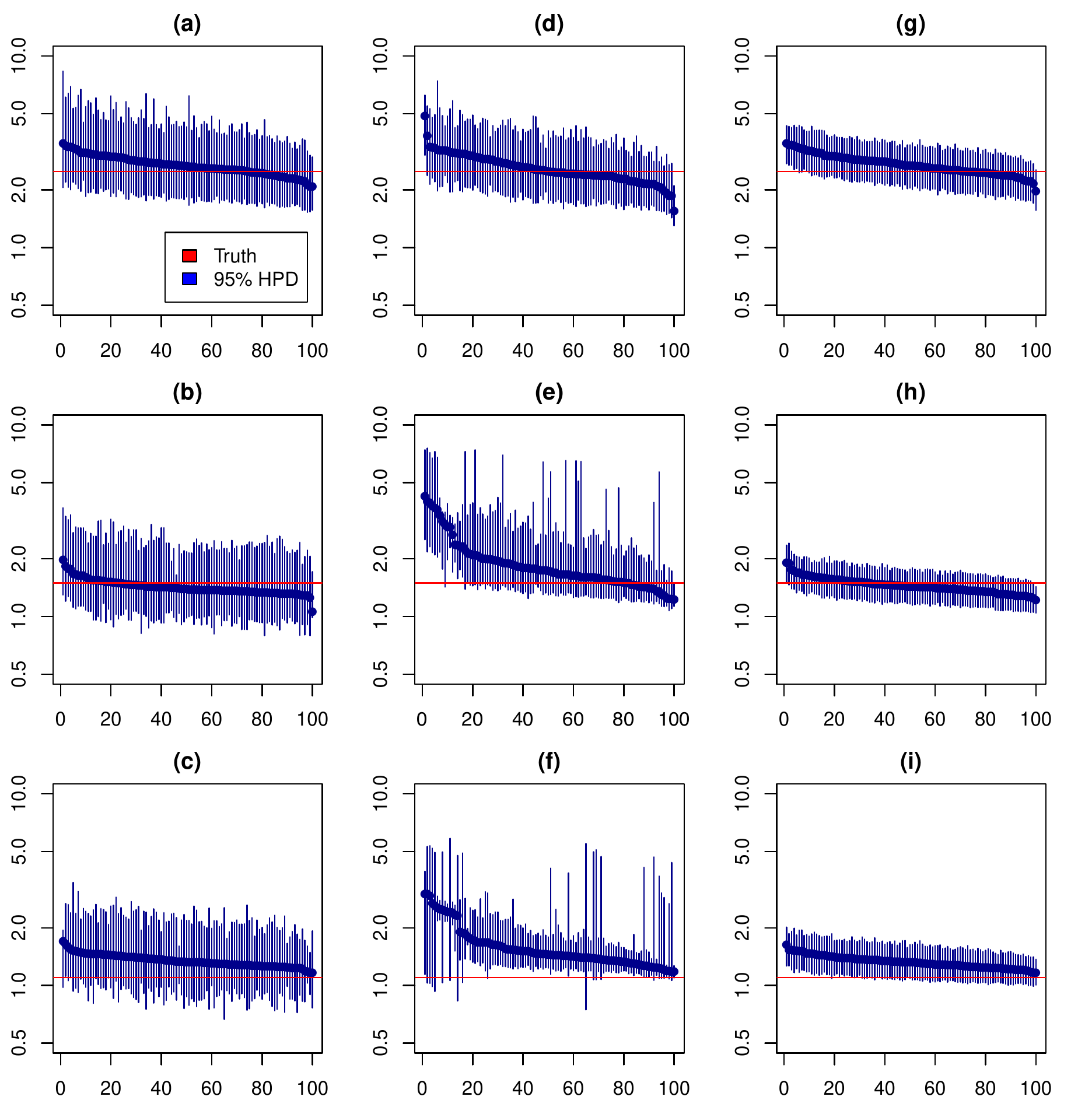}
\caption{Estimates of $R_{(0)}$ from true stochastic SIR trees using inference methods by column, with 
	\stochCoalSIR{} (a, b, c), \deterCoalSIR{} (d, e, f), and \BDSIR{} (g, h, i).  The truth varies 
	by row, with $R_{0}=2.4975$ (a, d, g), $R_{0}=1.4970$ (b, e, h), and $R_{0}=1.0978$ (c, f, i).
}
\end{mainFigure}
\begin{mainFigure}[H]
    \centering
    	\includegraphics[width=3.5in]{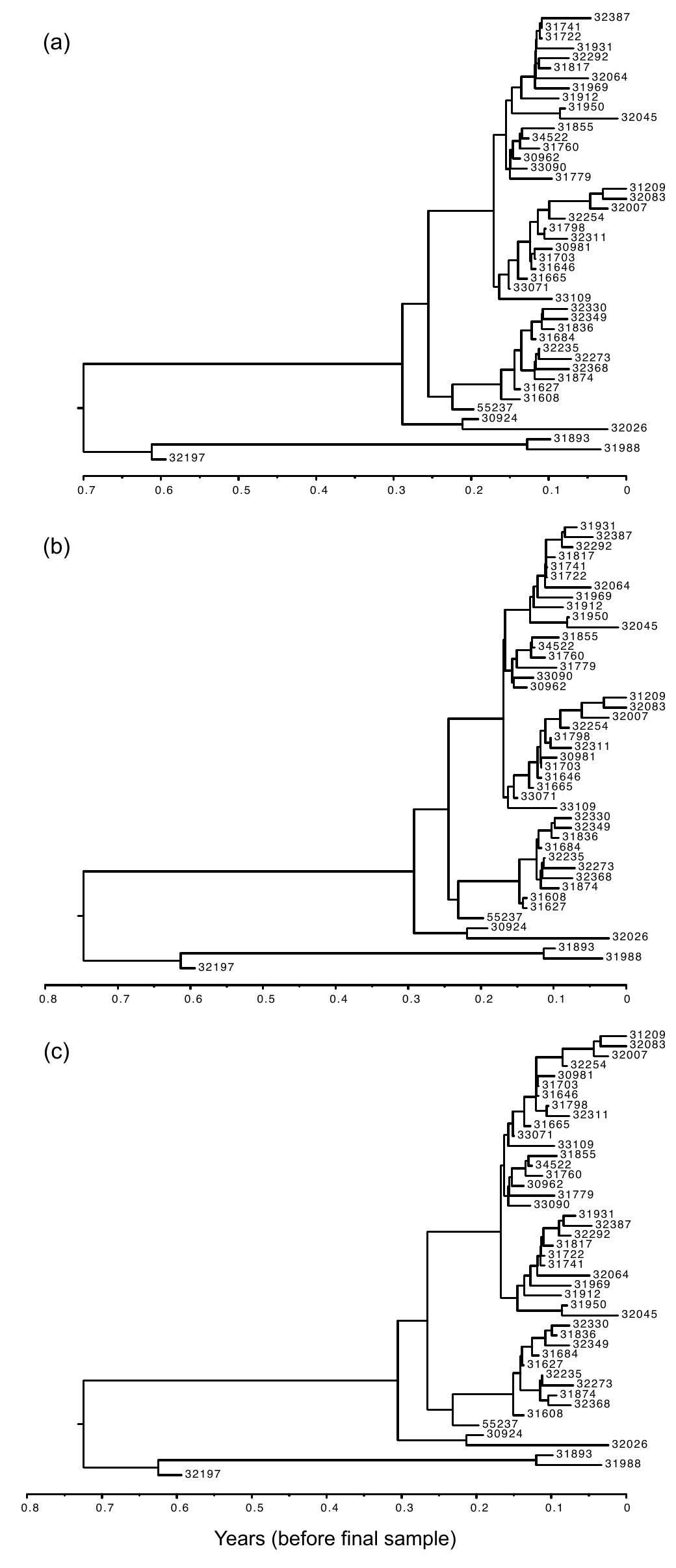}
\caption{Representative influenza A (H1N1) posterior trees from inference using the (a) \BDSIR{}, (b) \stochCoalSIR{}, and (c) \deterCoalSIR{} inference models.
}
\end{mainFigure}
\begin{mainTable}[H]
\begin{center}
\caption{\large{Results for Simulated Sequences:  $R_{0} \approx 2.50$, $S_{0}=999$}}
\label{table:simSeq}
\begin{tabular}{|c|c|c|c|c|c|c|c|c|}
\hline
$\eta$ & Inference & Truth & Mean & Median & Error & Bias & Relative & 95\% HPD \\ 
&  &  &  &  &  &  &  HPD width & accuracy \\ 
	\hline
	\hline
& Stoch.Coal.SIR & 2.50 & 2.41 & 2.16 & 0.13 & -0.11 & 0.97 & 97.00\% \\
$\mathcal{R}_0$ & Deter.Coal.SIR & 2.50 & 2.78 & 2.03 & 0.38 & 0.05 & 0.79 & 87.00\% \\
& BDSIR & 2.50 & 3.21 & 2.84 & 0.15 & 0.14 & 1.86 & 100.00\% \\
   \hline
   \hline 
& Stoch.Coal.SIR & 0.30 & 0.16 & 0.13 & 0.52 & -0.52 & 0.82 & 47.00\% \\
$\gamma$ & Deter.Coal.SIR & 0.30 & 0.25 & 0.16 & 0.56 & -0.28 & 0.97 & 56.00\% \\
& BDSIR & 0.30 & 0.17 & 0.14 & 0.52 & -0.52 & 1.13 & 84.00\% \\
   \hline
   \hline
& Stoch.Coal.SIR & 999 & 1805 & 1148 & 0.32 & 0.21 & 5.12 & 99.00\% \\
$S_{(0)}$ & Deter.Coal.SIR & 999 & 2384 & 1565 & 0.66 & 0.60 & 6.54 & 100.00\% \\
& BDSIR & 999 & 4002 & 2611 & 1.70 & 1.70 & 10.38 & 99.00\% \\
   \hline
   \hline
& Stoch.Coal.SIR & (varies) & 51.67 & 48.89 & 0.26 & 0.23 & 0.61 & 37.00\% \\
$z_{(0)}$ & Deter.Coal.SIR & (varies) & 49.13 & 46.46 & 0.22 & 0.20 & 0.26 & 29.00\% \\
& BDSIR & (varies) & 31.16 & 29.52 & 0.51 & 0.51 & 0.79 & 18.00\% \\
   \hline
\end{tabular}
\end{center}
 \end{mainTable}
\begin{mainTable}[H]
\small
\begin{center}
\caption{
\large{Simulation Study Results for Fixed Trees:  $R_{0}\approx2.50$ and $S_{0}=999$, 
$R_{0}\approx1.50$ and $S_{0}=499$, $R_{0}\approx1.10$ and $S_{0}=499$}}
\begin{tabular}{|c|c|c|c|c|c|c|c|c|}
\hline
$\eta$ & Inference & Truth & Mean & Median & Error & Bias & Relative & 95\% HPD \\ 
&  &  &  &  &  &  &  HPD width & accuracy \\ 
	\hline
	\hline
& Stoch.Coal.SIR & 2.50 & 2.84 & 2.68 & 0.12 & 0.09 & 0.98 & 100.00\% \\
$\mathcal{R}_0$ & Deter.Coal.SIR & 2.50 & 2.68 & 2.49 & 0.13 & 0.04 & 0.81 & 98.00\% \\
& \BDSIR{} & 2.50 & 2.73 & 2.67 & 0.12 & 0.08 & 0.55 & 94.00\%\\ 
   \hline
   \hline 
& Stoch.Coal.SIR & 0.30 & 0.27 & 0.25 & 0.19 & -0.13 & 1.14 & 99.00\% \\
$\gamma$ & Deter.Coal.SIR & 0.30 & 0.32 & 0.29 & 0.16 & 3.14\mbox{\sc{e}-3} & 1.27 & 99.00\% \\
& \BDSIR{} & 0.30 & 0.28 & 0.27 & 0.13 & -0.09 & 0.62 & 95.00\%\\ 
   \hline
   \hline
& Stoch.Coal.SIR & 999 & 1390 & 921 & 0.19 & -0.03 & 3.85 & 100.00\% \\
$S_{(0)}$ & Deter.Coal.SIR & 999 & 1807 & 1133 & 0.52 & 0.29 & 4.59 & 98.00\% \\
& \BDSIR{} & 999 & 1591 & 1142 & 0.39 & 0.24 & 3.42 & 99.00\% \\ 
   \hline
   \hline
& Stoch.Coal.SIR & (varies) & 41.81 & 40.35 & 0.03 & 0.01 & 0.20 & 99.00\% \\
$z_{(0)}$ & Deter.Coal.SIR & (varies) & 41.17 & 39.99 & 0.03 & 0.01 & 0.07 & 76.00\% \\
& \BDSIR{} & (varies) & 40.89 & 39.72 & 8.65\mbox{\sc{e}-4} & -5.13\mbox{\sc{e}-4} & 3.43\mbox{\sc{e}-3} & 97.00\% \\  
   \hline
	\hline
	\hline
	\hline
& Stoch.Coal.SIR & 1.50 & 1.48 & 1.37 & 0.09 & -0.06 & 0.81 & 100.00\% \\
$\mathcal{R}_0$ & Deter.Coal.SIR & 1.50 & 1.80 & 1.49 & 0.24 & 0.15 & 0.52 & 85.00\% \\
& \BDSIR{} & 1.50 & 1.46 & 1.43 & 0.08 & -0.03 & 0.47 & 99.00\% \\ 
   \hline
   \hline 
& Stoch.Coal.SIR & 0.30 & 0.19 & 0.17 & 0.40 & -0.40 & 1.06 & 85.00\% \\
$\gamma$ & Deter.Coal.SIR  & 0.30 & 0.26 & 0.23 & 0.27 & -0.22 & 1.15 & 89.00\% \\
& \BDSIR{} & 0.30 & 0.26 & 0.25 & 0.18 & -0.18 & 0.72 & 97.00\% \\ 
   \hline
   \hline
& Stoch.Coal.SIR & 499 & 599 & 390 & 0.25 & -0.22 & 3.56 & 100.00\% \\
$S_{(0)}$ & Deter.Coal.SIR & 499 & 562 & 361 & 0.44 & -0.26 & 3.36 & 91.00\% \\
& \BDSIR{} & 499 & 996 & 714 & 0.51 & 0.49 & 4.63 & 100.00\% \\ 
   \hline
   \hline
& Stoch.Coal.SIR & (varies) & 76.47 & 68.24 & 0.55 & 0.54 & 0.58 & 99.00\% \\
$z_{(0)}$ & Deter.Coal.SIR & (varies) & 91.03 & 72.51 & 0.39 & 0.38 & 0.42 & 88.00\% \\
& \BDSIR{} & (varies) & 69.11 & 66.51 & 0.34 & -0.31 & 0.20 & 94.00\% \\  
    \hline
	\hline
	\hline
	\hline
 & Stoch.Coal.SIR & 1.10 & 1.39 & 1.32 & 0.22 & 0.22 & 1.09 & 99.00\% \\
$\mathcal{R}_0$ & Deter.Coal.SIR & 1.10 & 1.68 & 1.44 & 0.46 & 0.46 & 0.59 & 25.00\% \\
 & BDSIR & 1.10 & 1.34 & 1.32 & 0.20 & 0.20 & 0.51 & 75.00\% \\
   \hline
   \hline 
 & Stoch.Coal.SIR & 0.25 & 0.17 & 0.15 & 0.37 & -0.36 & 1.11 & 84.00\% \\
$\gamma$ & Deter.Coal.SIR & 0.25 & 0.22 & 0.18 & 0.30 & -0.22 & 1.16 & 86.00\% \\
 & BDSIR & 0.25 & 0.28 & 0.26 & 0.12 & 0.09 & 0.92 & 100.00\% \\
   \hline
   \hline
 & Stoch.Coal.SIR & 499 & 608 & 398 & 0.24 & -0.18 & 3.38 & 100.00\% \\
$S_{(0)}$ & Deter.Coal.SIR & 499 & 553 & 337 & 0.42 & -0.26 & 3.08 & 92.00\% \\
 & BDSIR & 499 & 1471 & 1040 & 1.21 & 1.21 & 6.52 & 99.00\% \\
   \hline
   \hline
 & Stoch.Coal.SIR & (varies) & 91.60 & 84.55 & 0.06 & 0.02 & 0.60 & 97.00\% \\
$z_{(0)}$ & Deter.Coal.SIR & (varies) & 112.79 & 90.37 & 0.26 & 0.26 & 0.94 & 85.00\% \\
 & BDSIR & (varies) & 82.98 & 80.93 & 0.02 & -0.01 & 0.08 & 88.00\% \\
   \hline
\end{tabular}
\end{center}
{}
\label{table:sim}
\end{mainTable}
\begin{mainTable}[H]
\footnotesize
\begin{center}
\caption{
\large{Epidemic Parameter Inference from H1N1 Sequences in New Zealand}}
\vspace{5mm}
\label{table:H1N1}
\begin{tabular}{|c|c|c|c|c|c|}
  \hline
Inference Model & $R_0$ & $\gamma$ & $S_0$ & Root of & Origin $z_{0}$ of the \\ 
& & & & the tree (yr) & epidemic (yr) \\
   \hline
   \hline
    & & & & &\\
\bf{\StochCoalSIR} & 1.46 & 27.08 & 6.90\mbox{\sc{e}4} & 0.53 & 0.69 \\
 & (1.04 - 2.14) & (4.20 - 64.03) & (175 - 2.86\mbox{\sc{e}5}) & (0.44 - 0.61) & (0.45 - 1.03) \\
& & & & &\\ 
\bf{\DeterCoalSIR} & 1.35 & 34.50 & 1.20\mbox{\sc{e}5} & 0.54 & 0.73 \\
 & (1.05 - 1.84) & (3.86 - 82.16) & (29 - 4.59\mbox{\sc{e}5}) & (0.45 - 0.62) & (0.47 - 1.04) \\
& & & & &\\ 
\bf{\BDSIR{}} & 1.61 & 27.72 & 2.22\mbox{\sc{e}4} & 0.49 & 0.53 \\ 
 & (1.09 - 2.29) & (6.82 - 55.04) & (259 - 9.38\mbox{\sc{e}4}) & (0.41 - 0.56) & (0.43 - 0.65) \\ 
   \hline
\end{tabular}
\end{center}
\vspace{3mm}
{Mean estimates (and 95\% HPD intervals) of each epidemic parameter inferred from seasonal influenza A (H1N1)
sequence data collected in the Canterbury region of New Zealand throughout the 2001 flu season.}
\end{mainTable}
\begin{mainTable}[H]
\begin{center}
\caption{
\large{Bayesian Prior Distributions}}
\label{table:priors}
\begin{tabular}{|c|c|c|c|c|c|}
\hline
& $R_0$ & $\gamma$ & $S_{(0)}$ & $z_{(0)}$ & $\psi/(\psi+\mu)$ \\
  \hline
   & & & & & \\
${R_0}\approx2.5$, ${S_0}=999$ & LogN(1, 1) & LogN(-1, 1) & LogN(7, 1) & Unif(0, 100) & Beta(1,1) \\
   & & & & & \\
   \hline
   & & & & & \\
${R_0}\approx1.5$, ${S_0}=499$ & LogN(0.5, 1) & LogN(-1, 1) & LogN(6, 1) & Unif(0, 500) & Beta(1,1) \\
   & & & & & \\
   \hline
   & & & & & \\
${R_0}\approx1.1$, ${S_0}=499$ & LogN(0.1, 1) & LogN(-1.5, 1) & LogN(6, 1) & Unif(0, 500) & Beta(1,1) \\
   & & & & & \\
   \hline
   & & & & & \\
* ${R_0}\approx1.1$, ${S_0}=999$ & LogN(0.1, 1) & LogN(-1.5, 1) & LogN(7, 1) & Unif(0, 500) & -- \\
   & & & & & \\
   \hline
   & & & & & \\
* ${R_0}\approx1.1$, ${S_0}=1999$ & LogN(0.1, 1) & LogN(-1.5, 1) & LogN(7.5, 1) & Unif(0, 500) & -- \\
   & & & & & \\
   \hline
   & & & & & \\
* ${R_0}\approx1.2$, ${S_0}=499$ & LogN(0.2, 1) & LogN(-1, 1) & LogN(6, 1) & Unif(0, 500) & -- \\
   & & & & & \\
   \hline
   & & & & & \\
H1N1 & Unif(0, 10) & LogN(3, 0.75) & LogN(13, 2) & Unif(0, 10) & Beta(1,1) \\
   & & & & & \\
HIV-1 & LogN(1, 1) & LogN(-1, 1) & LogN(7, 1) & Unif(0, 100) & Beta(1,1) \\
   & & & & & \\
   \hline
\end{tabular}
\end{center}
{Prior distributions for the re-estimation of SIR parameters -- the reproductive ratio 
$R_0$, the rate of removal $\gamma$, the number of susceptible individuals at the start of 
the epidemic $S_{(0)}$, the time of origin $z_{(0)}$, and the sampling proportion $\psi/(\psi+\mu)$ for \BDSIR{} -- from the simulated trees, seasonal influenza A (H1N1), and human immunodeficiency virus (HIV-1) data analyses. 
LogN($M$, $S$) is a log-normal distribution with mean $M$ and 
standard deviation $S$ in log space.  *Only applies to deterministic coalescent SIR, see details in the supporting information.}
\end{mainTable}

\clearpage

\begin{center}
\Large{\bf{Supporting material for ``Inferring epidemiological dynamics with Bayesian coalescent inference:  The merits of deterministic and stochastic models''}}
\end{center}

\section{Sampling from the prior}

In order to assess the correctness of our implementation of the
deterministic coalescent SIR and stochastic coalescent SIR models, for
each model we used the MCMC algorithm to sample trees from the
corresponding distribution $f(\tree|\eta)$, and compared these samples
with coalescent trees simulated directly under the model.

The chosen $\eta$ included $\beta=7.5\times 10^{-4}$, $\gamma=0.3$,
$S_0=999$ and $z_0=30$. The comparisons were performed for trees
generated from 20 leaves, sampled at integer times 0 through 19,
inclusive.

For the deterministic coalescent SIR model, the direct simulation
involved numerically solving the Eqs.~(1)--(3) in the main text for
$t\in[0,30]$ and using this solution in combination with Eq.~(10)
in the main text to determine the instantaneous coalescent rate
$\lambda(\tau)$. This rate was used to simulate each of the coalescent
trees in the usual fashion for heterochronous leaf times.  In the case
that the MRCA was not reached before the origin time of the epidemic,
the tree was discarded and the simulation repeated.

The direct simulation proceeded in a similar way for the stochastic
coalescent SIR model, the major difference being that the
stochasticity of this model required each coalescent tree to be
simulated under a distinct realization of the stochastic trajectory.

Comparisons between the direct simulation and MCMC results are shown
in Figures \ref{fig:detCoalValidation} and
\ref{fig:stochCoalValidation} for three different summary
statistics and show very close agreement.

\vspace{3 mm}

\section{Validation through simulated data analysis}

As part of the validation of our implementation of the two coalescent SIR models, trees were 
simulated by their own methods (using stochastically- and deterministically-generated SIR trajectories, 
as discussed in the Methods section of the main paper), and relevant epidemiological parameters were inferred using the 
stochastic and deterministic coalescent SIR models.  Tables 1 and 2 show the results of these analyses, 
indicative of correct implementations. 

Analyses for varying $R_0$ (and necessarily, slightly varied other parameters, such as the birth rate $\beta$) are provided 
in Tables S3 and S4.  Results from tests of the influence of broader priors (with larger standard deviations in log space) are shown in Table S4.  
It appears that allowance of broader priors reduces 95\% HPD coverage in some cases (e.g., for parameter $R_0$) when using the \deterCoalSIR{} inference model, as they increase error and bias.

Finally, it was noticed that even for the higher true parameter values of $R_{0}=2.50$ and $S_{0}=999$, under which \deterCoalSIR{} is expected to 
perform relatively well, there was an inability to accurately estimate the origin parameter $z_0$.  Figure S3 provides some insight into 
this conundrum by examining the trajectories used for tree simulation and subsequent analysis. 

\section{H1N1 data selection}

Initially, the H1N1 dataset contained 45 sequences.  The ages of the inferred trees (Figure S4) using the original 45 sequences 
extended more than 1.5 years into the past for each of the SIR models, which is contrary to what we expect for a single, current strain of seasonal influenza.  
Three taxa (labelled 32197, 31893, and 31988) were hypothesized to belong to a unique strain, e.g., an additional seeding from outside the 
Canterbury region or a low-lying previous strain.  Removing these three taxa caused the inferred trees to behave as expected, i.e., tree heights and 
epidemic origin $z_0$ less than a year old.  It also raised the estimated $R_0$ values for all three SIR models (initially 1.24, 1.10, and 1.55 for \stochCoalSIR{}, 
\deterCoalSIR{}, and \BDSIR{}, respectively), as well as those for $\gamma$ (initially 8.74, 12.65, and 11.33 for \stochCoalSIR{}, 
\deterCoalSIR{}, and \BDSIR{}, respectively).  

It will be interesting to further investigate the interplay between influenza strains and its contribution to the overall dynamics.  
For the closed SIR models discussed in this manuscript, however, this additional complexity leads to increased chance of model misspecification and misleading results.  
Therefore, we focused our attention on the analyses using 42 sequences.

\section{HIV-1 data analysis}

The original HIV-1 dataset \citep{Hue:2005} was agglomerated from both acute and chronic 
infections sampled in the United Kingdom (UK) and constitutes six phylogenetic clusters, 
from which the five used here (Clusters 1-4 and 6) were drawn.  These particular clusters, with the omission of Cluster 5, were chosen simply 
for the purpose of direct comparison with \cite{Kuhnert:2014}.  Our extension to the 
models allowed us to imprint respective tip dates on the sequence data, sampled from 1999 
to 2003, for inclusion in the likelihood computation. 

For the selected five clusters, the nucleotide alignments 
contained 41, 62, 29, 26, and 35 sequences, respectively, each with 952 sites.  The substitution scheme 
chosen for phylogenetic analysis was the symmetric and independent general time reversible 
model (GTR), with gamma distributed rate variation and explicit proportion of invariable sites (GTR+G+I).  Following \cite{Hue:2005}, the substitution rate 
was set to $\num{2.55}$\mbox{\sc{e}-4} substitutions per site per year.  All other parameters were estimated conjointly, 
and the Bayesian prior distributions are presented in Table 4:  Bayesian prior distributions.

The pathophysiology of HIV is multifarious, and the patterns of its advancement within an 
infected host change throughout time.  In addition to increased complexity potentially caused by recombination events, the transition between HIV's acute and 
chronic phases alters the host's infectivity \citep{Guss:1994}.  The SIR compartmental 
model used for this particular phylodynamic analysis on the UK cluster data does not 
allow for independent infection rates for the acute and chronic phases (but see \cite{VolzFrost:2012} and \cite{Volz:2013}).   However, 
in this study we did not attempt to estimate the infection rate $\beta$ and thus did not 
expect such a difference to significantly impact the estimation of the parameters of 
interest:  the basic reproductive number $R_0$, removal rate $\gamma$, size of the initial susceptible population 
$S_{0}$, and origin of the outbreak $z_{0}$.

\section{HIV-1 inference results}  

In regard to parameter inference from the serially-sampled HIV-1 sequence data, the \stochCoalSIR{}, \deterCoalSIR{}, and \BDSIR{} methods were most alike in 
light of the $R_{0}$ results.  The medians and HPD intervals for all clusters pertaining to this parameter,
(especially Clusters 1, 2, 3, and 6), were very close, and those of Cluster 4 were still congruent across the three analyses (Figure S5).  

The coalescent SIR models and \BDSIR{} disagreed with respect to the age of the most recent common ancestor and the origin $z_0$ (Figure S6).  The coalescent SIR models also exhibited much larger 95\% HPD intervals for $z_0$ in each of the clusters; 
while \BDSIR{} encompassed an average of 16 years, the \stochCoalSIR{} and \deterCoalSIR{} models had averages of 49 and 37 years, respectively.   
Furthermore, the estimated age of the common ancestor of the tree was older under the coalescent SIR
models than the estimates reported by either \BDSIR{} or the original data analysis \citep{Hue:2005} for each cluster.  This was also true for the time of origin for the epidemic, 
although for certain clusters the differences between the coalescent estimates of the origin $z_0$ and the birth-death estimates were much greater than others (e.g., Cluster 3).   

The estimates of removal rate $\gamma$ 
from Clusters 1 and 6 were very similar across the three methods (Figure S7).  However, both coalescent SIR
models estimated considerably higher $\gamma$ values for Clusters 2-4 than BDSIR.  This is reflective of the simulation study results, where the two coalescent models 
did not perform as well as \BDSIR{} for the removal parameter.

Median estimates for the initial susceptible population $S_0$ were quite similar in all methods for Clusters 1-4, although \BDSIR{} displayed much wider HPD intervals than \stochCoalSIR{} and \deterCoalSIR{} (Figure S8).
In Cluster 6, the coalescent SIR models showed the smallest HPD intervals for their individual analyses on each cluster, while the opposite was true for \BDSIR{}.  
There was also a disparity between the median estimates for the two coalescent approaches and that of \BDSIR{} for Cluster 6. 
To this effect, it should be noted that the 
number of infections accrued throughout the duration of the epidemic was reported as $N_{e}=1,350$ by Hu{\'e} \textit{et al.}  
This casts some suspicion on the low susceptible population estimates obtained by the \stochCoalSIR{} and 
\deterCoalSIR{} methods (median estimates of $S_{0}=727$ and $S_{0}=693$, respectively), since they appear lower than the estimated number of infected individuals from the original study.

There is disagreement in the literature in regard to the modelling of HIV-1 evolutionary dynamics under stochastic or deterministic processes \citep{Nijhuis:1998,Rouzine:1999,Achaz:2004,Shriner:2004}.
The predicament dwells in the observation that the actual effective population size $N_e$ for HIV-1 is often smaller 
than the total population size \citep{Kouyos:2006}.
While most of this debate has focused on within-host population dynamics, many of the arguments hold when considering the broader epidemic dynamics of host-to-host transmission.
As previously mentioned, the  appropriateness of these descriptions is hinged on the magnitude of the infected population, precisely, the effective infected population size. Consequently, even when the total infected population is quite large there may yet be significant stochastic effects in play.

Finally, as mentioned in the main article, the existence of two distinct infectious stages and the possibility of large effects due to recombination are reasons for any discrepancy produced by these SIR inference models.

\section{Example XML} 

Below is an example XML for simulating 100 trees and trajectories in MASTER \citep{Vaughan:MASTER}.  This example is for $R_0=2.4975$ and $S_0=999$.  
The simulation ends when the infected $I$ population returns to zero, i.e., when the last infected individual is removed.

\lstset{
  basicstyle=\ttfamily,
  columns=fullflexible,
  showstringspaces=false,
  commentstyle=\color{gray}\upshape
}

\lstdefinelanguage{XML}
{
  morestring=[b]",
  morestring=[s]{>}{<},
  morecomment=[s]{<?}{?>},
  stringstyle=\color{black},
  identifierstyle=\color{darkblue},
  keywordstyle=\color{cyan},
  morekeywords={xmlns,version,type}
}

\footnotesize{
\begin{lstlisting}

<beast version=`2.0' 
namespace=`master.beast:beast.core.parameter:beast.evolution.tree.TreeHeightLogger'>

    <run spec=`InheritanceEnsemble'
	 nTraj=`100'
	 samplePopulationSizes=`true'
	 verbosity=`1'>

        <model spec=`InheritanceModel' id=`model'>
            <population spec=`Population' id=`S' populationName=`S'/>
            <population spec=`Population' id=`I' populationName=`I'/>
            <population spec=`Population' id=`R' populationName=`R'/>
            <population spec=`Population' id=`Rh' populationName=`Rh'/>
            
            <!-- infection reaction -->
            <reaction spec=`InheritanceReaction' reactionName=`Infection' rate=`0.00075'>
                S + I -> 2I
            </reaction>
            
            <!-- recovery reaction -->
            <reaction spec=`InheritanceReaction' reactionName=`Recovery' rate=`0.25'>
                I -> R
            </reaction>
            
            <!-- sampling reaction -->
            <reaction spec=`InheritanceReaction' reactionName=`Sampling' rate=`0.05'>
                I -> Rh
            </reaction>
        </model>

        <initialState spec=`InitState'>
            <populationSize spec=`PopulationSize' population=`@S' size=`999'/>
            <lineageSeed spec=`Individual' population=`@I'/>
        </initialState>

	<populationEndCondition spec=`PopulationEndCondition'
				population=`@I'
				threshold=`0'
				exceedCondition=`false'/>

	<inheritancePostProcessor spec=`LineageFilter'
				  reactionName=`Sampling'
				  discard=`false'/>
		
        <output spec=`NewickOutput' fileName=`SIR.newick'/>
        <output spec=`NexusOutput' fileName=`SIR.nexus'/>
        <output spec=`JsonOutput' fileName=`SIR.json'/>

    </run>
</beast>
\end{lstlisting}}

\beginsupplement
\begin{suppFigure}
    \begin{center}
      \includegraphics[width=0.4\textwidth]{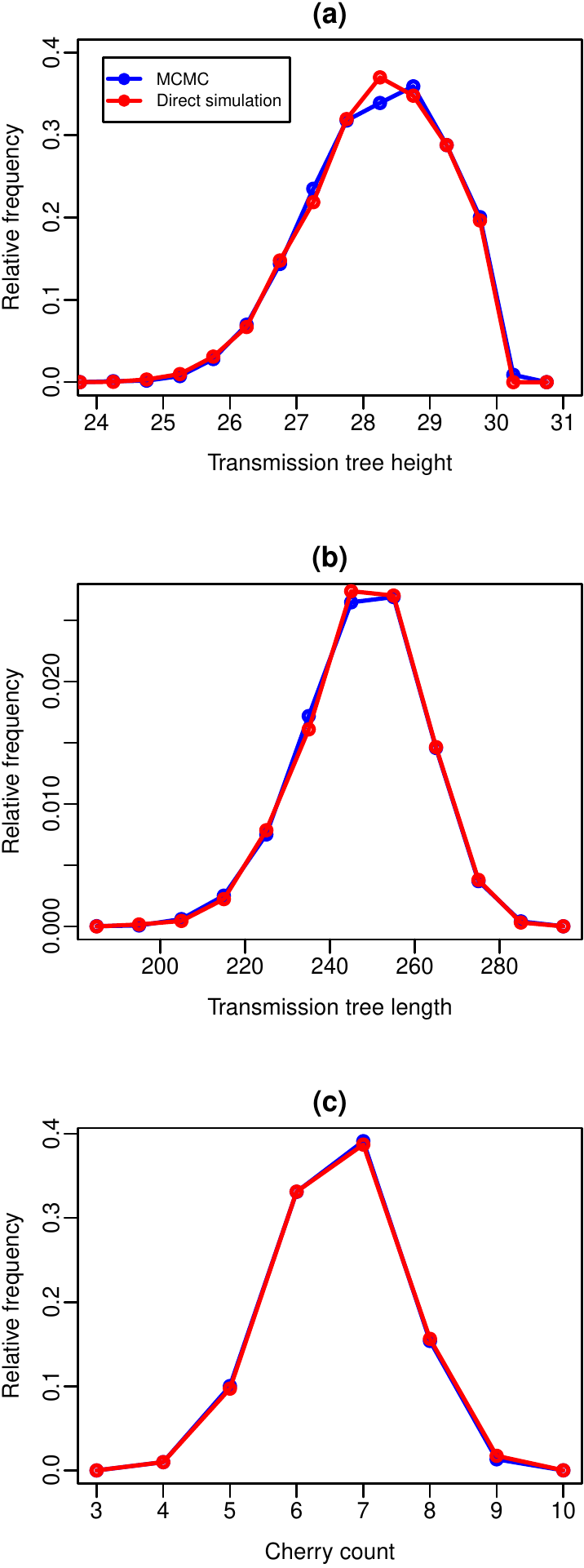}
    \end{center}
    \caption{Comparison between distributions of summary statistics of
      trees sampled using MCMC employing our implementation of the
      \emph{deterministic coalescent SIR model} likelihood and those
      calculated, and those of trees sampled using direct
      simulation. Summary statistics shown are (a) the age of the MRCA
      of the transmission tree, (b) the sum of all edge lengths in the
      tree, and (c) the total number of two-leaf clades in the tree.}
    \label{fig:detCoalValidation}
\end{suppFigure}
\begin{suppFigure}
    \begin{center}
      \includegraphics[width=0.4\textwidth]{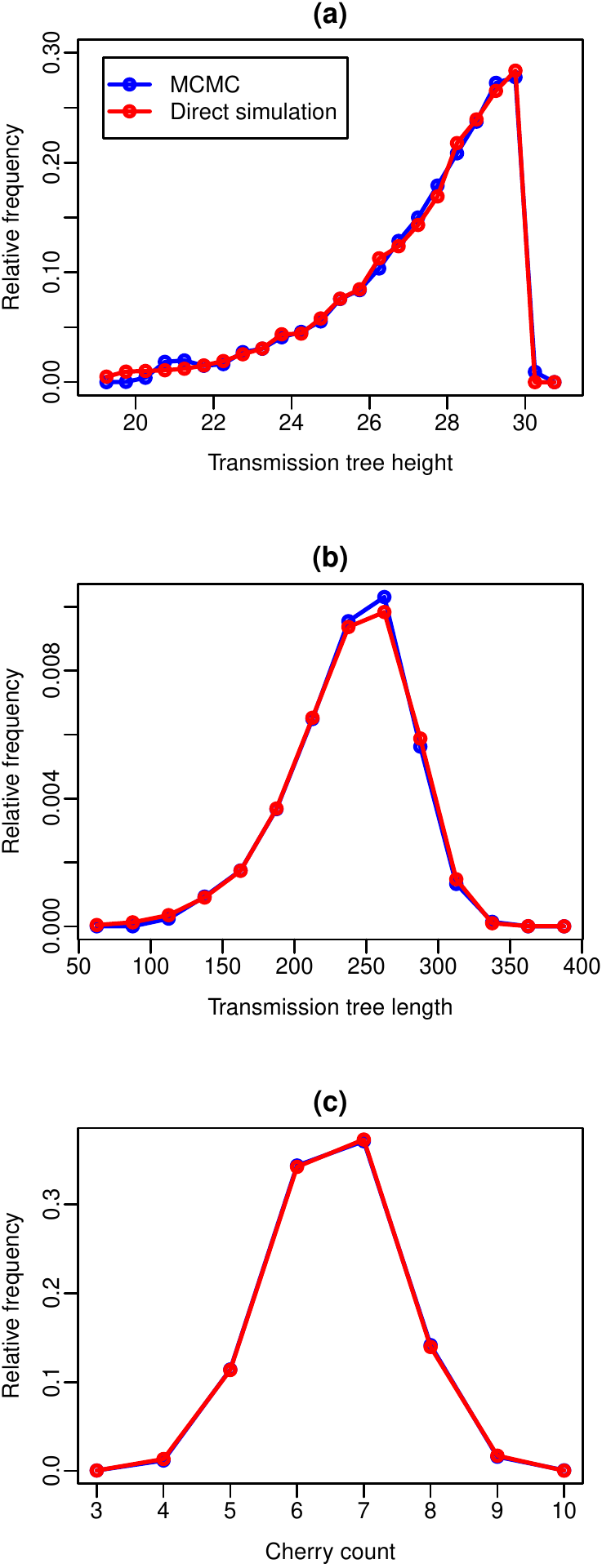}
    \end{center}
    \caption{Comparison between distributions of summary statistics of
      trees sampled using MCMC employing our implementation of the
      \emph{stochastic coalescent SIR model} likelihood and those
      calculated, and those of trees sampled using direct
      simulation. Summary statistics shown are (a) the age of the MRCA
      of the transmission tree, (b) the sum of all edge lengths in the
      tree, and (c) the total number of two-leaf clades in the tree.}
    \label{fig:stochCoalValidation}
\end{suppFigure}

\begin{suppFigure}
  \includegraphics[width=\textwidth]{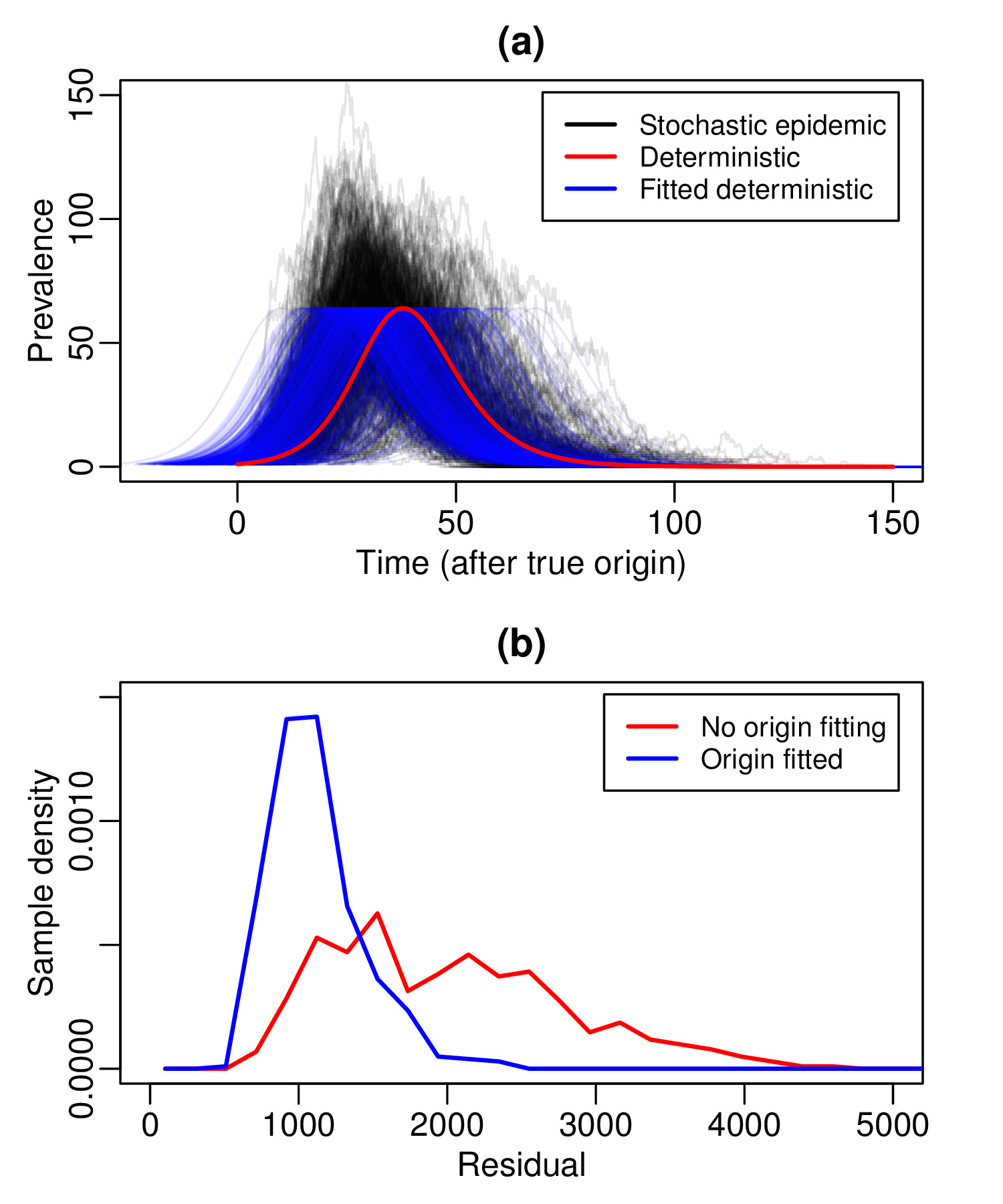}
  \caption{(a) True stochastic SIR trajectories simulated jointly alongside phylogenies, with 
  the corresponding trajectories used by \deterCoalSIR{}.  Adjusting \deterCoalSIR{} 
  to fit the underlying stochastic trajectories causes major shifts to the origin $z_0$.  
  (b) Deterministic residuals with $z_0$ either fitted or not.}
  \label{fig:originFit}
\end{suppFigure}
\begin{suppFigure}[ht]
    \centering
{%
    	\includegraphics[width=2.25in]{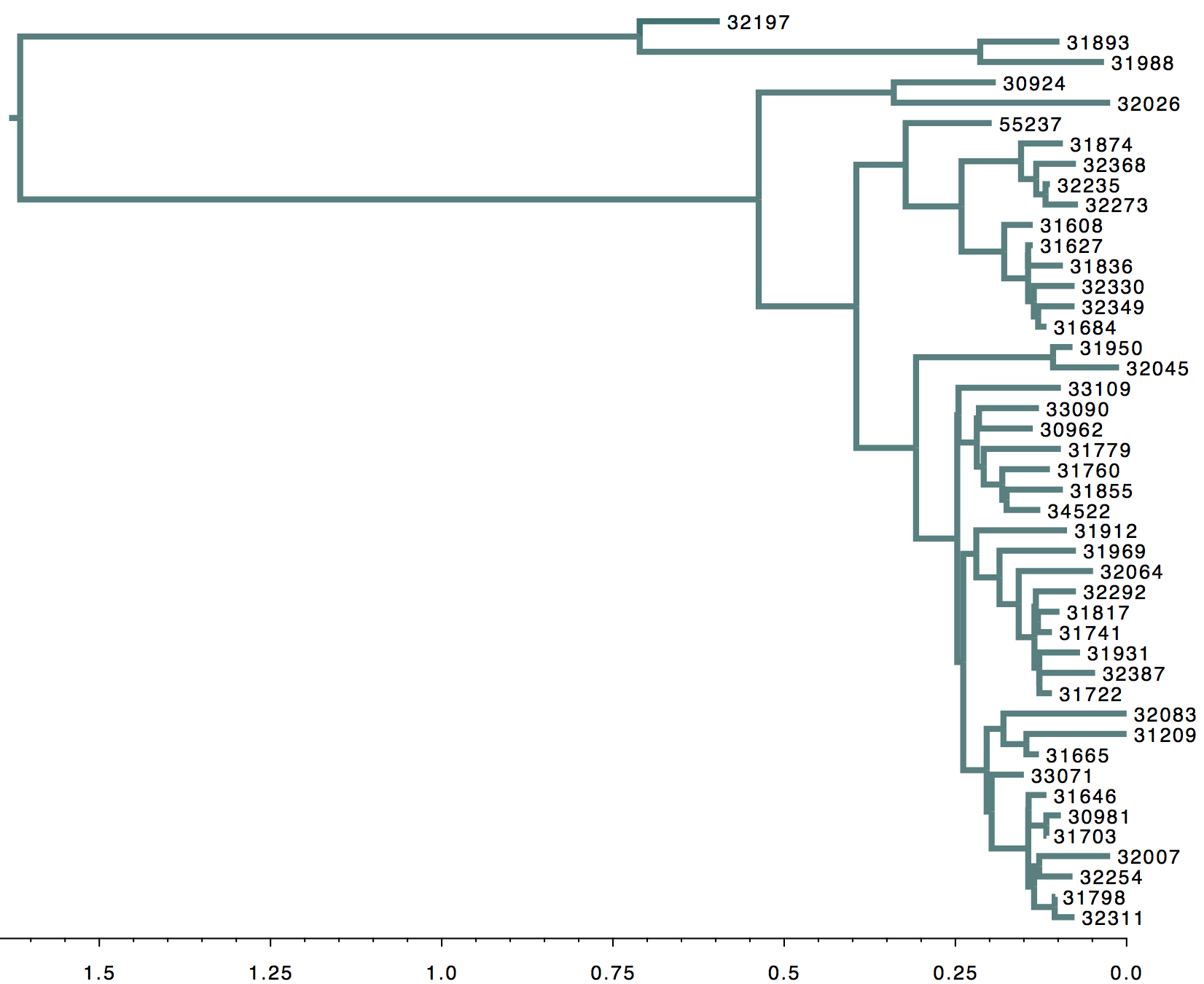}
}
\quad
{%
    	\includegraphics[width=2.25in]{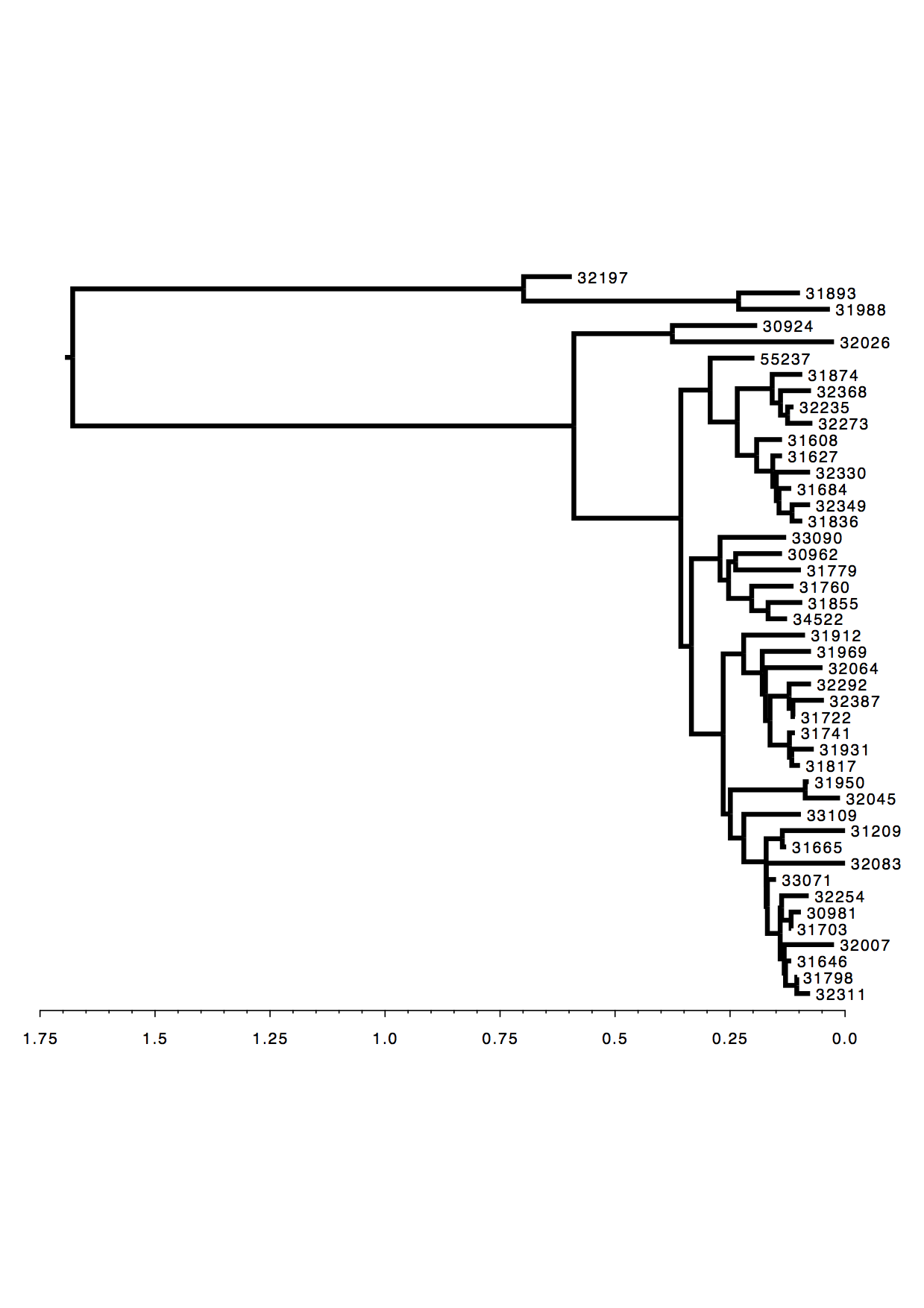}
}
\quad
{%
    	\includegraphics[width=2.25in]{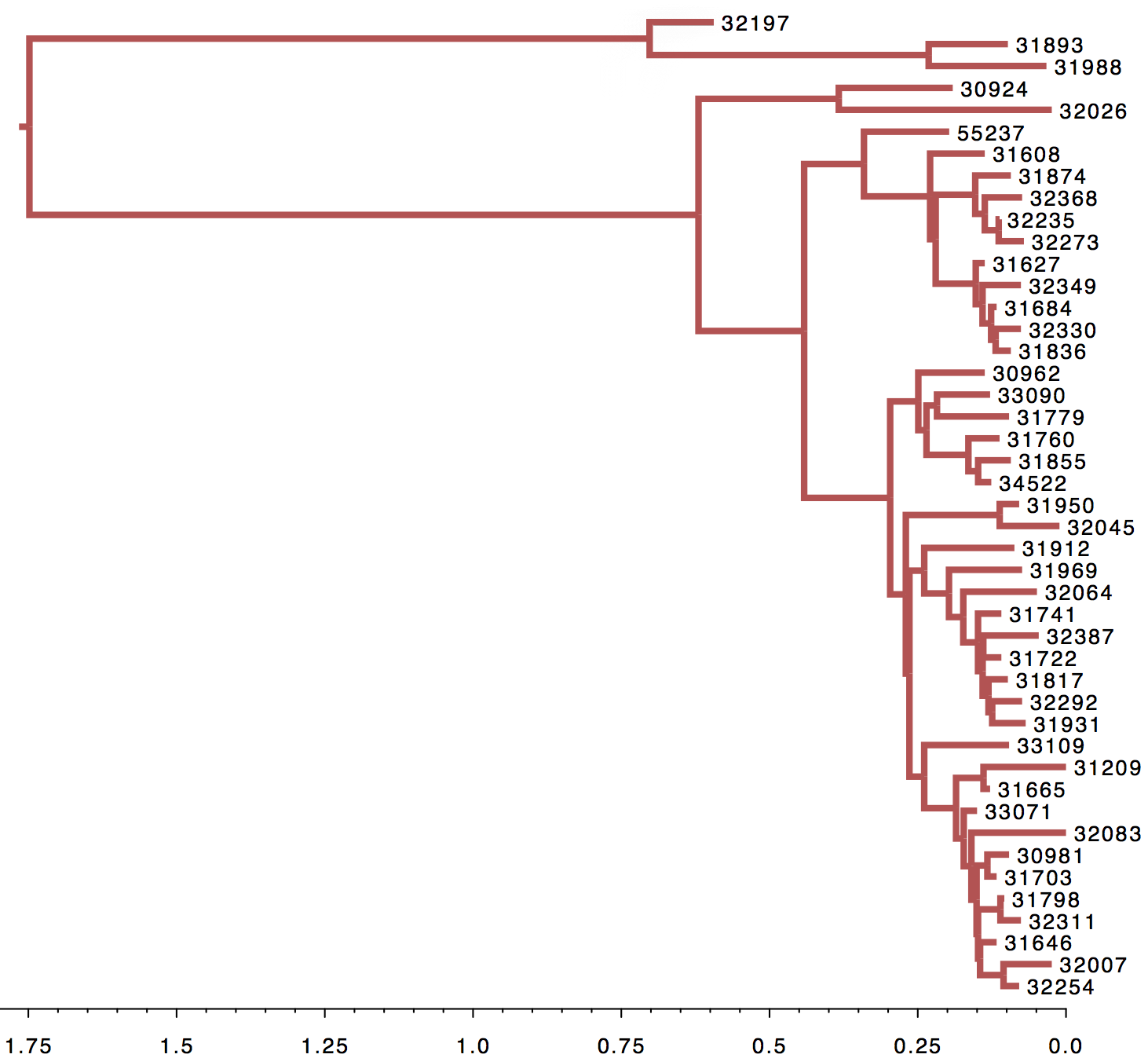}
}
\caption{Representative influenza A (H1N1) posterior trees from inference using the \stochCoalSIR{} (left), \deterCoalSIR{} (right), \BDSIR{} (bottom) models.
}
\end{suppFigure}
\begin{suppFigure}[!ht]
\begin{center}
\includegraphics[width=\textwidth]{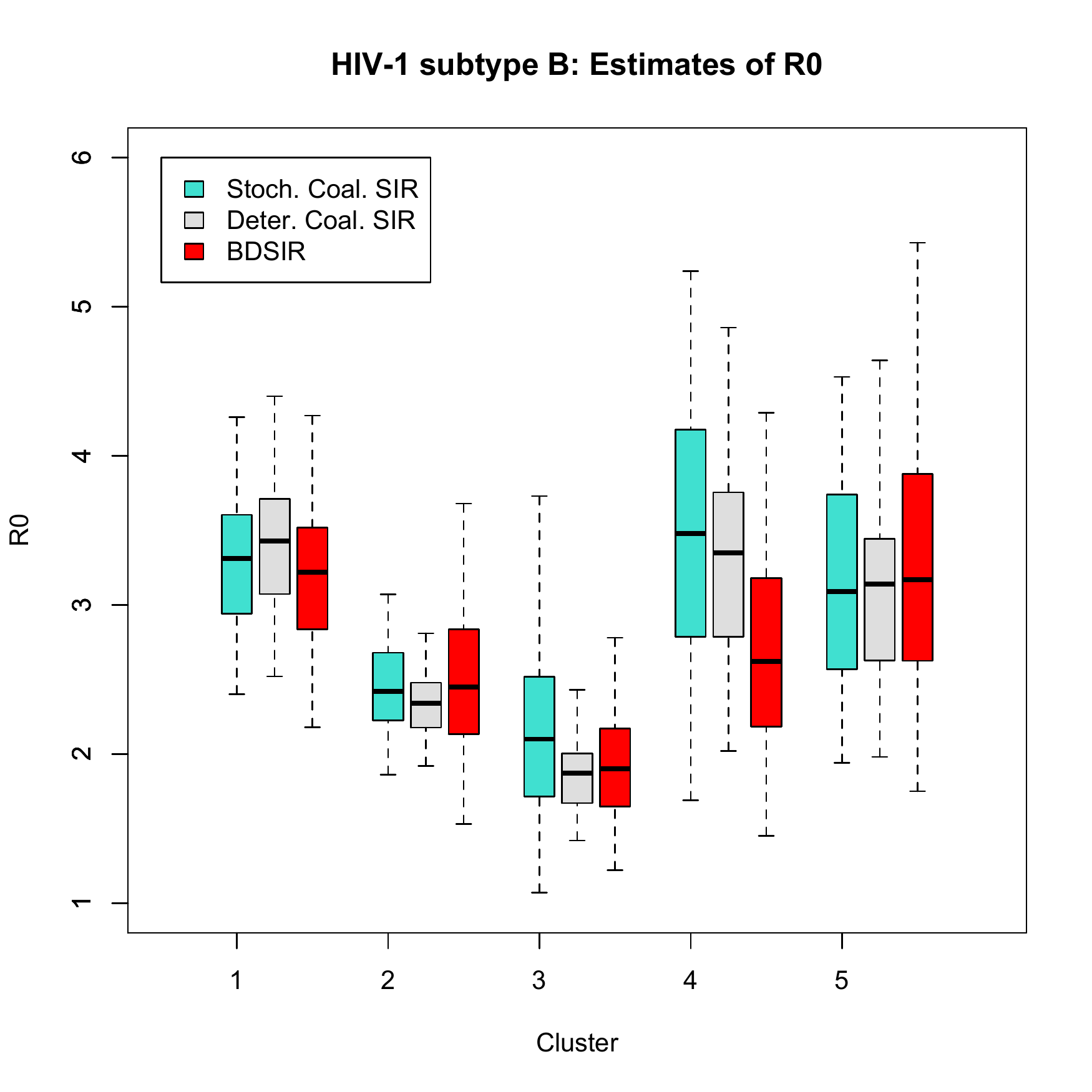}
\end{center}
\caption{
95\% HPD intervals of $R_0$ for the HIV-1 subtype B UK cluster 
analyses using coalescent and birth-death methods.}
\label{fig:HIV_R0}
\end{suppFigure}
\begin{suppFigure}[!ht]
\begin{center}
\includegraphics[width=\textwidth]{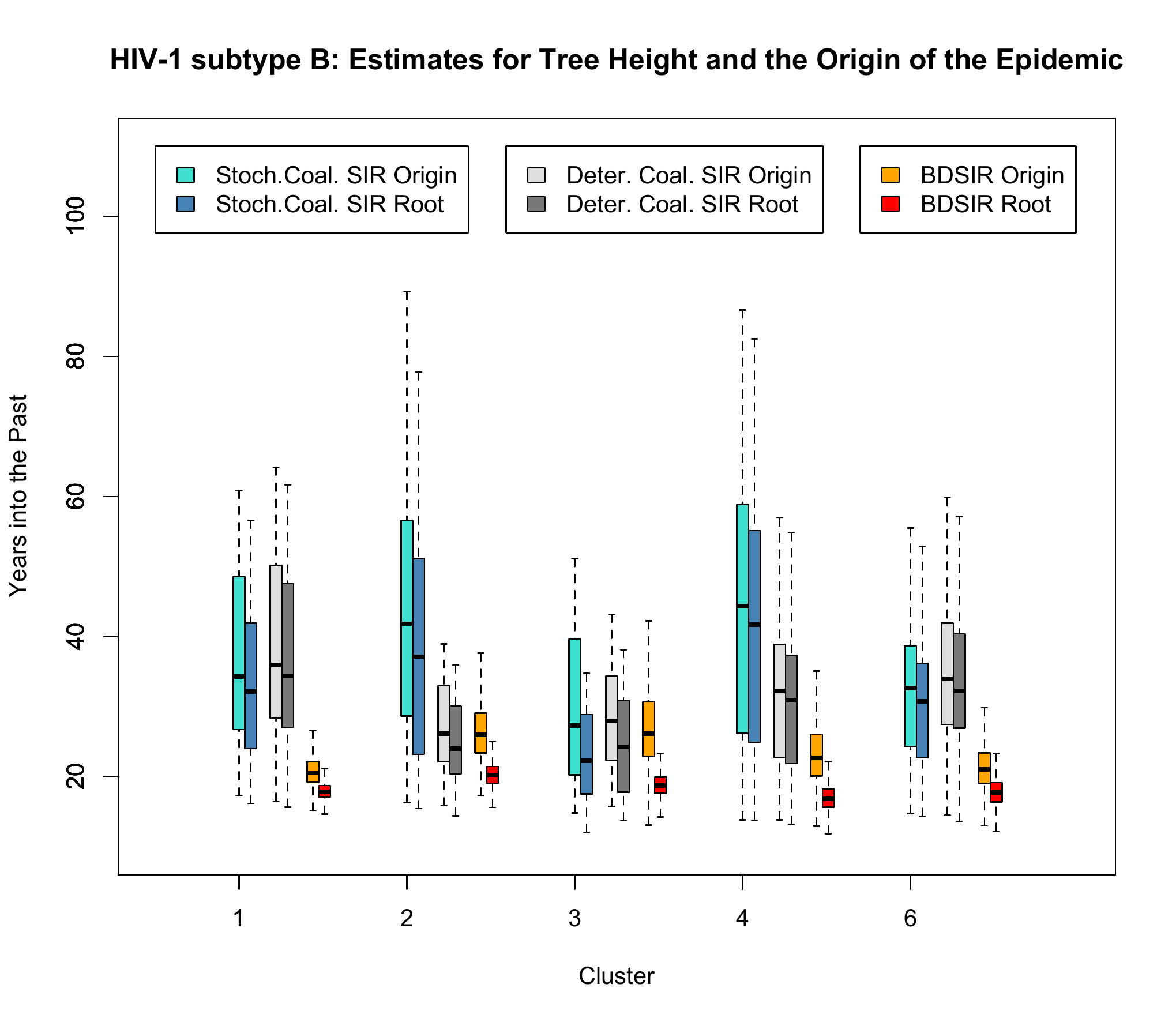}
\end{center}
\caption{
95\% HPD intervals for coalescent [\protect\cite{Volz:2012}] and birth-death [\protect\cite{Kuhnert:2014}]
estimations of the time into the past at which the root of the HIV-1 tree and introduction 
of the first infection occurred.}
\label{fig:HIV_HeightandOrigin}
\end{suppFigure}
\begin{suppFigure}[!ht]
\begin{center}
\includegraphics[width=\textwidth]{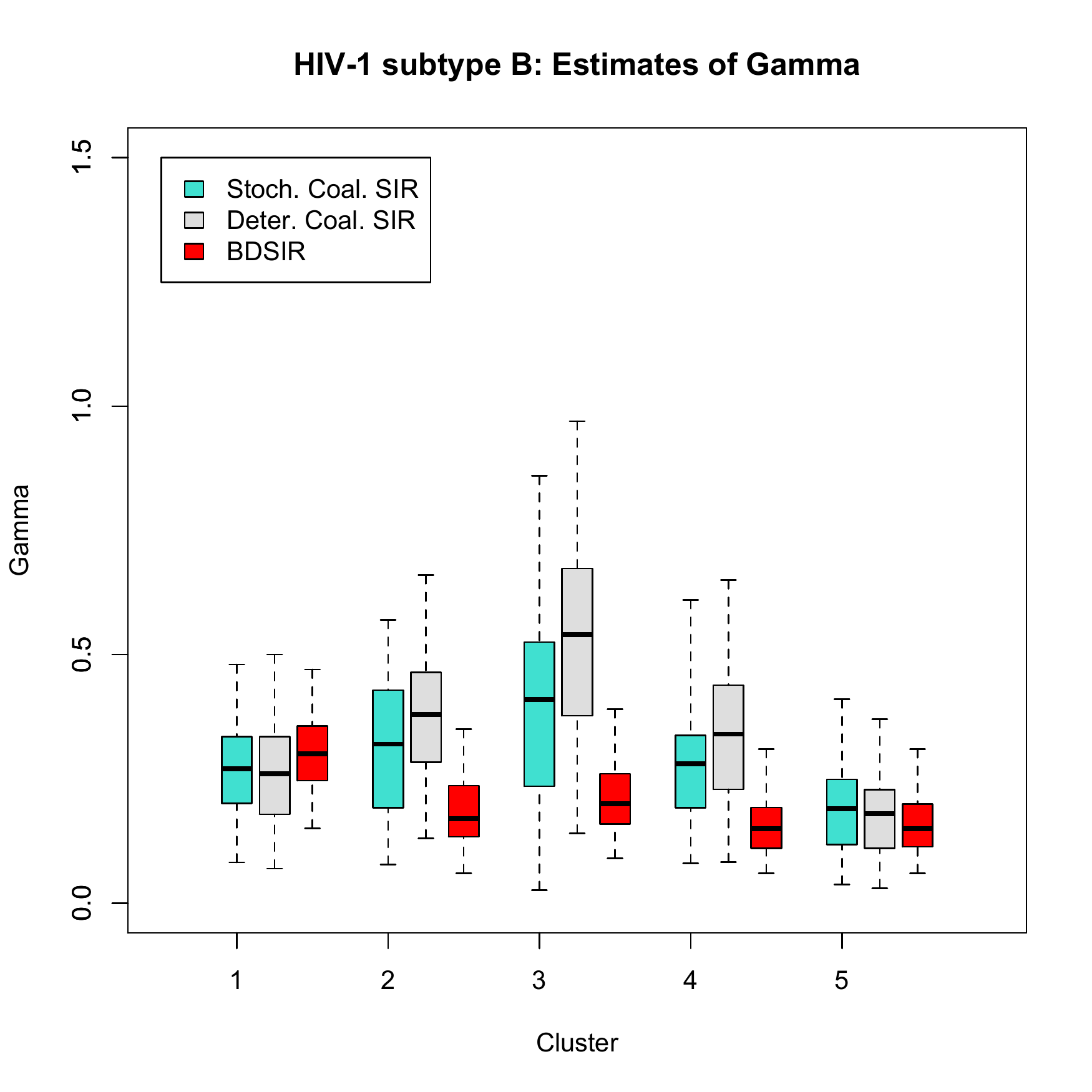}
\end{center}
\caption{
95\% HPD intervals of $\gamma$ for the HIV-1 subtype B UK cluster 
analyses using coalescent and birth-death methods.}
\label{fig:HIV_gamma}
\end{suppFigure}
\begin{suppFigure}[!ht]
\begin{center}
\includegraphics[width=\textwidth]{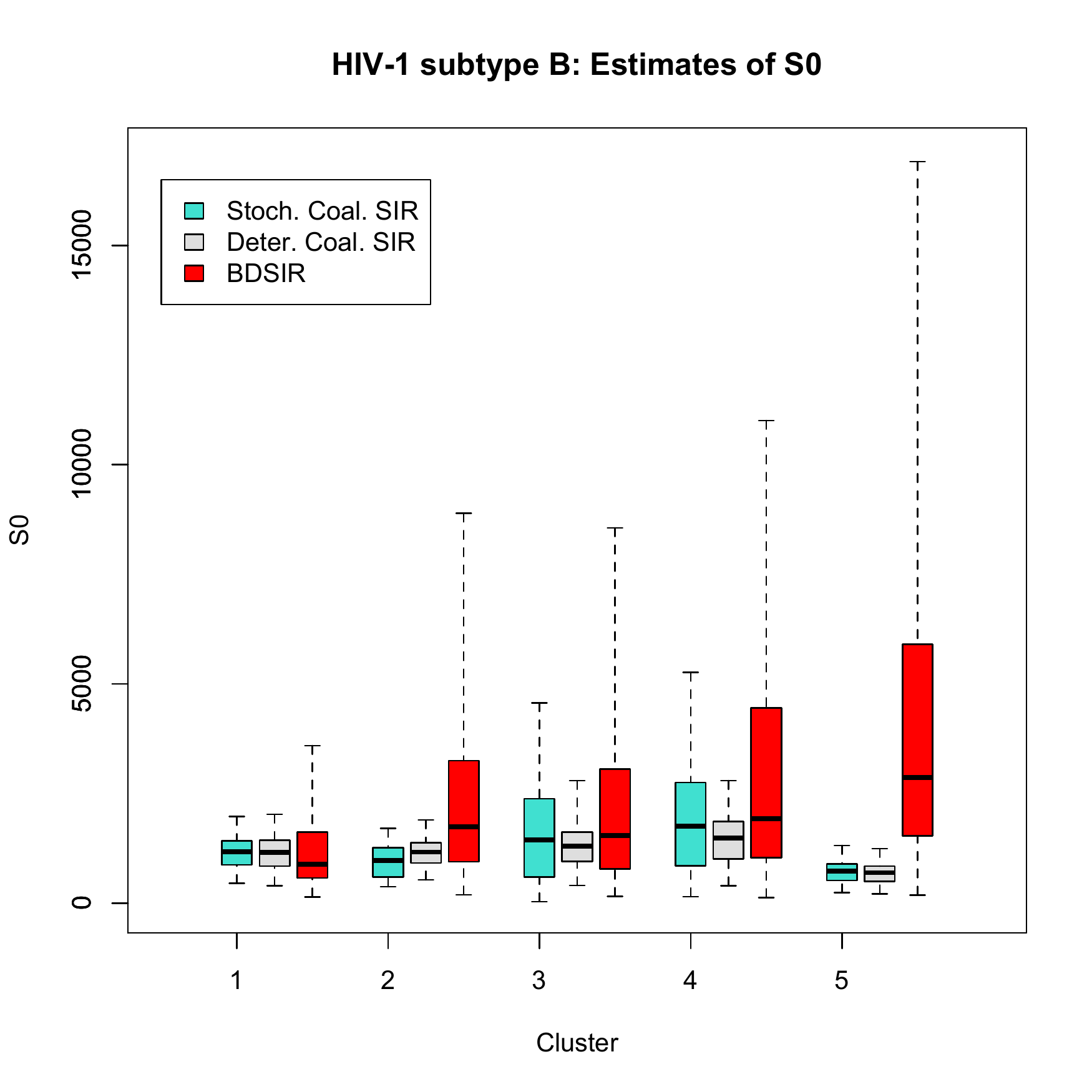}
\end{center}
\caption{95\% HPD intervals of $S_0$ for the HIV-1 subtype B UK cluster 
analyses using coalescent and birth-death methods.}
\label{fig:HIV_S0}
\end{suppFigure}
\begin{suppTable}[!ht]
\begin{center}
\caption{\large{Simulation Study Results for Stochastic Coalescent Trees}}
\begin{tabular}{|c|c|c|c|c|c|c|c|c|}
\hline
$\eta$ & Inference & Truth & Mean & Median & Error & Bias & Relative & 95\% HPD \\ 
&  &  &  &  &  &  &  HPD width & accuracy \\ 
	\hline
	\hline
$\mathcal{R}_0$ & Stoch.Coal.SIR & 2.50 & 2.81 & 2.64 & 0.11 & 0.08 & 0.95 & 100.00\% \\
& Deter.Coal.SIR & 2.50 & 2.73 & 2.65 & 0.14 & 0.06 & 0.85 & 96.00\% \\
   \hline
   \hline 
$\gamma$ & Stoch.Coal.SIR & 0.30 & 0.28 & 0.26 & 0.16 & -0.11 & 1.17 & 99.00\% \\
& Deter.Coal.SIR & 0.30 & 0.30 & 0.28 & 0.18 & -0.03 & 1.20 & 99.00\% \\
   \hline
   \hline
$S_{(0)}$ & Stoch.Coal.SIR & 999 & 1456 & 986 & 0.21 & 0.02 & 3.93 & 100.00\% \\
& Deter.Coal.SIR & 999 & 1720 & 1057 & 0.48 & 0.24 & 4.28 & 99.00\% \\
   \hline
   \hline
$z_{(0)}$ & Stoch.Coal.SIR & (varies) & 42.36 & 40.43 & 0.03 & 0.02 & 0.20 & 98.00\% \\
& Deter.Coal.SIR & (varies) & 41.25 & 39.77 & 0.03 & 0.01 & 0.07 & 64.00\% \\ 
   \hline
\end{tabular}
\end{center}
\label{table:simStochCoalTrees}
 \end{suppTable}
\begin{suppTable}[!ht]
\begin{center}
\caption{\large{Simulation Study Results for Deterministic Coalescent Trees}}
\begin{tabular}{|c|c|c|c|c|c|c|c|c|}
\hline
$\eta$ & Inference & Truth & Mean & Median & Error & Bias & Relative & 95\% HPD \\ 
&  &  &  &  &  &  &  HPD width & accuracy \\ 
	\hline
	\hline
$\mathcal{R}_0$ & Stoch.Coal.SIR & 2.50 & 2.44 & 2.37 & 0.06 & -0.05 & 0.67 & 100.00\% \\
& Deter.Coal.SIR & 2.50 & 2.51 & 2.46 & 0.08 & -0.01 & 0.59 & 99.00\% \\
   \hline
   \hline 
$\gamma$ & Stoch.Coal.SIR & 0.30 & 0.33 & 0.31 & 0.07 & 0.05 & 1.00 & 100.00\% \\
& Deter.Coal.SIR & 0.30 & 0.32 & 0.30 & 0.10 & 0.02 & 0.79 & 100.00\% \\
   \hline
   \hline
$S_{(0)}$ & Stoch.Coal.SIR & 999 & 1586 & 1142 & 0.26 & 0.20 & 3.83 & 100.00\% \\
& Deter.Coal.SIR & 999 & 1426 & 1030 & 0.36 & 0.13 & 3.03 & 100.00\% \\
   \hline
   \hline
$z_{(0)}$ & Stoch.Coal.SIR & 44.12 & 45.52 & 44.74 & 0.02 & 0.01 & 0.19 & 93.00\% \\
& Deter.Coal.SIR & 44.12 & 44.34 & 44.11 & 0.02 & 1.93\mbox{\sc{e}-3} & 0.08 & 92.00\% \\
   \hline
\end{tabular}
\end{center}
\label{table:simDetCoalTrees}
 \end{suppTable}
\begin{suppTable}[!ht]
\begin{center}
\caption{\large{Results for Homochronous Sampling}}
\begin{tabular}{|c|c|c|c|c|c|c|c|c|}
\hline
$\eta$ & Inference & Truth & Mean & Median & Error & Bias & Relative & 95\% HPD \\ 
&  &  &  &  &  &  &  HPD width & accuracy \\ 
	\hline
	\hline
$\mathcal{R}_0$ & Stoch.Coal.SIR & 2.50 & 3.04 & 2.74 & 0.13 & 0.11 & 1.32 & 100.00\% \\
& Deter.Coal.SIR & 2.50 & 4.05 & 3.29 & 0.34 & 0.32 & 2.38 & 100.00\% \\
& BDSIR & 2.50 & 2.84 & 2.49 & 0.16 & 0.03 & 1.45 & 97.00\% \\
   \hline
   \hline 
$\gamma$ & Stoch.Coal.SIR & 0.30 & 0.26 & 0.23 & 0.25 & -0.21 & 1.43 & 100.00\% \\
& Deter.Coal.SIR & 0.30 & 0.26 & 0.19 & 0.36 & -0.31 & 2.03 & 100.00\% \\
& BDSIR & 0.30 & 0.23 & 0.17 & 0.42 & -0.42 & 2.04 & 100.00\% \\
   \hline
   \hline
$S_{(0)}$ & Stoch.Coal.SIR & 999 & 1660 & 1065 & 0.18 & 0.09 & 4.75 & 100.00\% \\
& Deter.Coal.SIR & 999 & 4127 & 679 & 0.78 & 0.09 & 10.24 & 100.00\% \\
& BDSIR & 999 & 1907 & 1320 & 0.41 & 0.41 & 4.86 & 100.00\% \\
   \hline
   \hline
$z_{(0)}$ & Stoch.Coal.SIR & 20.0 & 20.17 & 19.82 & 0.09 & -0.03 & 0.43 & 95.00\% \\
& Deter.Coal.SIR & 20.0 & 19.09 & 19.21 & 0.09 & -0.05 & 0.19 & 73.00\% \\
& BDSIR & 20.0 & 36.56 & 29.38 & 0.55 & 0.54 & 4.24 & 100.00\% \\
    \hline
\end{tabular}
\end{center}
\label{table:simDetCoalTrees}
 \end{suppTable}
\begin{suppTable}[!ht]
\begin{center}
\caption{
\large{Simulation Study Results: The Effect of Broader Priors on Deterministic Coalescent SIR}}
\label{table:simLowerS0broadPriors}
\end{center}
\begin{tabular}{|c|c|c|c|c|c|c|c|c|}
\hline
$\eta$ & St. Dev. & Truth & Mean & Median & Error & Bias & Relative & 95\% HPD \\ 
&  &  &  &  &  &  & HPD width & accuracy \\ 
	\hline
	\hline
$\mathcal{R}_0$ & 2 & 1.50 & 2.06 & 1.75 & 0.40 & 0.35 & 0.86 & 79.00\% \\
$\mathcal{R}_0$ & 1 & 1.50 & 1.80 & 1.49 & 0.24 & 0.15 & 0.52 & 85.00\% \\
$\mathcal{R}_0$ & 2 & 2.50 & 3.31 & 2.85 & 0.34 & 0.24 & 1.43 & 95.00\% \\
$\mathcal{R}_0$ & 1 & 2.50 & 2.68 & 2.49 & 0.13 & 0.04 & 0.80 & 99.00\% \\
   \hline
   \hline 
$\gamma$ & 2 & 0.30 & 0.31 & 0.23 & 0.37 & -0.12 & 1.59 & 96.00\% \\
$\gamma$ & 1 & 0.30 & 0.26 & 0.23 & 0.27 & -0.22 & 1.15 & 89.00\% \\
$\gamma$ & 2 & 0.30 & 0.31 & 0.25 & 0.33 & -0.09 & 1.59 & 95.00\% \\
$\gamma$ & 1 & 0.30 & 0.32 & 0.29 & 0.16 & 3.14\mbox{\sc{e}-3} & 1.27 & 99.00\% \\
   \hline
   \hline
$S_{(0)}$ & 2 & 499 & 2041 & 249 & 1.40 & 0.49 & 7.75 & 85.00\% \\
$S_{(0)}$ & 1 & 499 & 562 & 361 & 0.44 & -0.26 & 3.36 & 91.00\% \\
$S_{(0)}$ & 2 & 999 & 3028 & 717 & 1.05 & 0.33 & 6.60 & 94.00\% \\
$S_{(0)}$ & 1 & 499 & 553.38 & 337 & 0.42 & -0.26 & 3.08 & 92.00\% \\
   \hline
   \hline
$z_{(0)}$ & 2 & (varies) & 65.10 & 62.01 & 0.04 & 0.03 & 0.25 & 86.00\% \\
$z_{(0)}$ & 1 & (varies) & 91.03 & 72.51 & 0.39 & 0.38 & 0.42 & 88.00\% \\
$z_{(0)}$ & 2 & (varies) & 40.97 & 39.85 & 0.03 & -6.78\mbox{\sc{e}-4} & 0.08 & 81.00\% \\
$z_{(0)}$ & 1 & (varies) & 112.79 & 90.37 & 0.26 & 0.26 & 0.94 & 85.00\% \\
   \hline
\end{tabular}
\end{suppTable}
\begin{suppTable}[!ht]
\begin{center}
\caption{\large{Comparison of Computation Times for Bayesian Inference of Epidemic Parameters from Genetic Sequence Data using SIR Models}}
\vspace{3mm}
\label{table:compTime}
\begin{tabular}{|c|c|c|}
\hline
Data Type & Inference Model & Mean time per million \\ 
 &  &  samples (MCMC) \\ 
	\hline
& Stoch.Coal.SIR & 20m 41s \\
Sim. Study ($R_{0} \approx 2.50$) & Deter.Coal.SIR & 3m 27s \\
& BDSIR & 56m 27s \\
   \hline 
& Stoch.Coal.SIR & 1h 43m 30s \\
Sim. Study ($R_{0} \approx 1.50$) & Deter.Coal.SIR & 3m 47s \\
& BDSIR & 41m 35s \\
   \hline
& Stoch.Coal.SIR & 1h 50m 41s \\
Sim. Study ($R_{0} \approx 1.10$) & Deter.Coal.SIR & 6m 45s \\
& BDSIR & 41m 21s \\
   \hline
& Stoch.Coal.SIR & 1h 20m 55s \\
H1N1 & Deter.Coal.SIR & 9m 44s \\
& BDSIR & 47m 33s \\
   \hline
& Stoch.Coal.SIR & 14h 37m 45s \\
HIV-1 & Deter.Coal.SIR & 7m 56s \\
& BDSIR & 1h 38m 54s \\
   \hline
\end{tabular}
\end{center}
\end{suppTable}
\begin{suppTable}[!ht]
\begin{center}
\caption{\large{Deterministic Coalescent SIR Results for Simulated Sequences:  
$R_{0}=1.0987$ and $S_{0}=499$, $R_{0}=1.0989$ and $S_{0}=999$, $R_{0}=1.09945$ and $S_{0}=1999$}}
\label{table:simSeq}
\begin{tabular}{|c|c|c|c|c|c|c|c|}
\hline
$\eta$ & Truth & Mean & Median & Error & Bias & Relative & 95\% HPD \\ 
&  &  &  &  &  &  HPD width & accuracy \\ 
	\hline
	\hline
$\mathcal{R}_0$ & $\approx$1.10 & 1.89 & 1.28 & 0.62 & 0.63 & 0.40 & 52.00\% \\
   \hline
$\gamma$ & 0.30 & 0.57 & 0.44 & 0.59 & 0.52 & 2.14 & 95.00\% \\
   \hline
$S_{(0)}$ & 499 & 1830 & 1222 & 1.50 & 1.31 & 11.49 & 96.00\% \\
   \hline
$z_{(0)}$ & (varies) & 109.55 & 76.21 & 0.61 & 0.54 & 0.35 & 37.00\% \\
	\hline
	\hline
	\hline
$\mathcal{R}_0$ & $\approx$1.10 & 1.55 & 1.35 & 0.25 & 0.25 & 0.45 & 16.00\% \\
   \hline
$\gamma$ & 0.30 & 0.27 & 0.24 & 0.20 & -0.12 & 1.23 & 61.00\% \\
   \hline
$S_{(0)}$ & 999 & 1293 & 804 & 0.27 & -0.10 & 3.58 & 64.00\% \\
   \hline
$z_{(0)}$ & (varies) & 117.39 & 99.75 & 0.25 & 0.18 & 0.23 & 23.00\% \\
	\hline
	\hline
	\hline
$\mathcal{R}_0$ & $\approx$1.10 & 1.37 & 1.22 & 0.16 & 0.16 & 0.28 & 40.00\% \\
   \hline
$\gamma$ & 0.30 & 0.26 & 0.24 & 0.18 & -0.14 & 1.13 & 64.00\% \\
   \hline
$S_{(0)}$ & 1999 & 2292 & 1531 & 0.23 & -0.18 & 3.45 & 66.00\% \\
   \hline
$z_{(0)}$ & (varies) & 150.39 & 138.69 & 0.23 & 0.20 & 0.32 & 18.00\% \\
   \hline
\end{tabular}
\end{center}
\end{suppTable}
\begin{suppTable}[!ht]
\begin{center}
\caption{\large{Simulation Study Details}}
\vspace{3mm}
\label{table:simSeq}
\begin{tabular}{|c|c|}
\hline
\bf{Type of simulated data} & \bf{Inference models used} \\ 
	\hline
	\hline
1. Varying $R_0$ and $S_0$ (orig.) &  \\
(a) $R_{0}\approx 1.1$, $S_{0}=499$, $\gamma=0.25$, $\psi=0.15$ & Deter.Coal.SIR, Stoch.Coal.SIR, BDSIR \\
(b) $R_{0}\approx 1.2$, $S_{0}=499$, $\gamma=0.30$, $\psi=0.15$ & Deter.Coal.SIR \\
(c) $R_{0}\approx 1.5$, $S_{0}=499$, $\gamma=0.30$, $\psi=0.15$ & Deter.Coal.SIR, Stoch.Coal.SIR, BDSIR \\
(d) $R_{0}\approx 1.5$, $S_{0}=999$, $\gamma=0.30$, $\psi=0.20$ & Deter.Coal.SIR, Stoch.Coal.SIR, BDSIR \\
(e) $R_{0}\approx 2.5$, $S_{0}=999$, $\gamma=0.30$, $\psi=0.05$ & Deter.Coal.SIR, Stoch.Coal.SIR, BDSIR \\
   \hline
   \hline
2. Varying $S_0$ for fixed $R_0$ &  \\
(a) $R_{0}\approx 1.1$, $S_{0}=499$, $\gamma=0.25$, $\psi=0.15$ & Deter.Coal.SIR, Stoch.Coal.SIR, BDSIR \\
(f) $R_{0}\approx 1.1$, $S_{0}=999$, $\gamma=0.30$, $\psi=0.20$ & Deter.Coal.SIR \\
(g) $R_{0}\approx 1.1$, $S_{0}=1999$, $\gamma=0.30$, $\psi=0.09$ & Deter.Coal.SIR \\
   \hline
   \hline
3. Contemporaneous sampling & \\
(d) $R_{0}\approx 1.5$, $S_{0}=999$, $\gamma=0.30$, $\psi=0.20$ & Deter.Coal.SIR, Stoch.Coal.SIR, BDSIR \\
(e) $R_{0}\approx 2.5$, $S_{0}=999$, $\gamma=0.30$, $\psi=0.05$ & Deter.Coal.SIR, Stoch.Coal.SIR, BDSIR \\
   \hline
   \hline
4. Phylogenetic uncertainty &  \\
(e) $R_{0}\approx 2.5$, $S_{0}=999$, $\gamma=0.30$, $\psi=0.05$ & Deter.Coal.SIR, Stoch.Coal.SIR, BDSIR \\	
   \hline
   \hline
5. Reparameterization (growth rate) &  \\
(e) $R_{0}\approx 2.5$, $S_{0}=999$, $\gamma=0.30$, $\psi=0.05$ & Deter.Coal.SIR \\
   \hline
\end{tabular}
\end{center}
\end{suppTable}
\begin{suppTable}[!ht]
\footnotesize
\begin{center}
\caption{
\large{Epidemic Parameter Estimations from HIV-1 Subtype B Sequence Data}}
\vspace{5mm}
\label{table:HIV}
\begin{tabular}{|c|c|c|c|c|c|}
  \hline
\uline{Inference Model} & $R_0$ & $\gamma$ & $S_0$ & Root of & Origin $z_{0}$ of the \\ 
HIV cluster & & & & the tree (yr) & epidemic (yr) \\
   \hline
   \hline
    & & & & &\\
\bf{\StochCoalSIR} & & & & &\\
------------------ & & & & & \\
Cluster 1 & 3.31 & 0.27 & 1165 & 1971 & 1969 \\ 
 & (2.40 - 4.26) & (8.17E-2 - 0.48) & (448 - 1974) & (1946-1987) & (1942-1986) \\
Cluster 2 & 2.42 & 0.32 & 976 & 1975 & 1972 \\ 
 & (1.86 - 3.07) & (7.72E-2 - 0.57) & (371 - 1701) & (1953 - 1988) & (1947 - 1988) \\
Cluster 3 & 2.10 & 0.41 & 1442 & 1979 & 1973 \\ 
 & (1.07 - 3.73) & (2.59\mbox{\sc{e}-2} - 0.86) & (33 - 4568) & (1959 - 1990) & (1943 - 1989) \\
Cluster 4 & 3.48 & 0.28 & 1757 & 1964 & 1961 \\ 
 & (1.69 - 5.24) & (0.08 - 0.61) & (148 - 5260) & (1922 - 1990) & (1918 - 1991) \\
Cluster 6 & 3.09 & 0.19 & 727 & 1972 & 1970 \\ 
 & (1.94 - 4.53) & (3.72\mbox{\sc{e}-2} - 0.41) & (236 - 1312) & (1950 - 1989) & (1947 - 1988) \\
   \hline
   \hline
   & & & & &\\
\bf{\DeterCoalSIR} & & & & &\\
-------------------- & & & & & \\
Cluster 1 & 3.43 & 0.26 & 1158 & 1969 & 1967 \\ 
 & (2.52 - 4.40) & (6.95E-2 - 0.50) & (397 - 2023) & (1941-1987) & (1939-1986) \\
Cluster 2 & 2.34 & 0.38 & 1163 & 1979 & 1977 \\ 
 & (1.92 - 2.81) & (0.13 - 0.66) & (530 - 1895) & (1967 - 1989) & (1964 - 1987) \\
Cluster 3 & 1.87 & 0.54 & 1298 & 1979 & 1975 \\ 
 & (1.42 - 2.43) & (0.14 - 0.97) & (399 - 2267) & (1965 - 1989) & (1960 - 1987) \\
Cluster 4 & 3.35 & 0.34 & 1479 & 1972 & 1971 \\ 
 & (2.02 - 4.86) & (8.22\mbox{\sc{e}-2} - 0.65) & (397 - 2792) & (1948 - 1990) & (1946 - 1989) \\
Cluster 6 & 3.14 & 0.18 & 693 & 1971 & 1969 \\ 
 & (1.98 - 4.64) & (2.99\mbox{\sc{e}-2} - 0.37) & (213 - 1241) & (1949 - 1989) & (1943 - 1988) \\
   \hline
   \hline
   & & & & &\\
\bf{\BDSIR{}} & & & & &\\ 
--------------- & & & & & \\
Cluster 1 & 3.22 & 0.30 & 880 & 1986 & 1983 \\ 
 & (2.18-4.27) & (0.15-0.47) & (142-3592) & (1983-1988) & (1978-1987) \\
Cluster 2 & 2.45 & 0.17 & 1745 & 1983 & 1978 \\ 
& (1.53-3.68) & (0.06-0.35) & (190-8892) & (1979-1986) & (1968-1984) \\ 
Cluster 3 & 1.90 & 0.20 & 1540 & 1985 & 1978 \\ 
 & (1.22-2.78) & (0.09-0.39) & (153-8558) & (1981-1988) & (1962-1986) \\ 
Cluster 4 & 2.62 & 0.15 & 1921 & 1987 & 1981 \\ 
 & (1.45-4.29) & (0.06-0.31) & (128-11007) & (1983-1990) & (1970-1988) \\ 
Cluster 6 & 3.17 & 0.15 & 2862 & 1986 & 1983 \\  
 & (1.73-5.43) & (0.06-0.31) & (183-16909) & (1981-1989) & (1975-1989) \\ 
   \hline
\end{tabular}
\end{center}
\end{suppTable}
\begin{suppTable}[!ht]
\small
\begin{center}
\caption{
\large{Deterministic Coalescent SIR Results from Trees Simulated with Higher $S_0$ (with Fixed $R_0$) 
and Higher $R_{0}$ (with Fixed $S_0$)}}
\vspace{3mm}
\begin{tabular}{|c|c|c|c|c|c|c|c|}
\hline
$\eta$ & Truth & Mean & Median & Error & Bias & Relative & 95\% HPD \\ 
&  &  &  &  &  &  HPD width & accuracy \\ 
	\hline
	\hline
$\mathcal{R}_0$ & 2.50 & 2.68 & 2.49 & 0.13 & 0.04 & 0.81 & 98.00\% \\
   \hline 
$\gamma$ & 0.30 & 0.32 & 0.29 & 0.16 & 3.14\mbox{\sc{e}-3} & 1.27 & 99.00\% \\
   \hline
$S_{(0)}$ & 999 & 1807 & 1133 & 0.52 & 0.29 & 4.59 & 98.00\% \\
   \hline
$z_{(0)}$ & (varies) & 41.17 & 39.99 & 0.03 & 0.01 & 0.07 & 76.00\% \\
	\hline
	\hline
$\mathcal{R}_0$ & 2.50 & 3.28 & 2.97 & 0.23 & 0.20 & 1.42 & 100.00\% \\
   \hline
$\gamma$ & 0.35 & 0.30 & 0.28 & 0.24 & -0.20 & 1.28 & 99.00\% \\
   \hline
$S_{(0)}$ & 4999 & 7733 & 4838 & 0.34 & 0.03 & 4.18 & 100.00\% \\
   \hline
$z_{(0)}$ & (varies) & 37.45 & 36.15 & 0.03 & 1.48e-3 & 0.06 & 56.00\% \\
	\hline
	\hline
$\mathcal{R}_0$ & 2.50 & 3.76 & 3.05 & 0.26 & 0.23 & 1.50 & 100.00\% \\
   \hline
$\gamma$ & 0.40 & 0.33 & 0.31 & 0.26 & -0.22 & 1.22 & 100.00\% \\
   \hline
$S_{(0)}$ & 9999 & 12,609 & 7405 & 0.35 & -0.15 & 3.31 & 100.00\% \\
   \hline
$z_{(0)}$ & (varies) & 34.99 & 34.28 & 0.04 & -1.94e-3 & 0.05 & 43.00\% \\
	\hline
	\hline
	\hline
$\eta$ & Truth & Mean & Median & Error & Bias & Relative & 95\% HPD \\ 
&  &  &  &  &  &  HPD width & accuracy \\ 
	\hline
	\hline
$\mathcal{R}_0$ & 2.50 & 2.68 & 2.49 & 0.13 & 0.04 & 0.81 & 98.00\% \\
   \hline 
$\gamma$ & 0.30 & 0.32 & 0.29 & 0.16 & 3.14\mbox{\sc{e}-3} & 1.27 & 99.00\% \\
   \hline
$S_{(0)}$ & 999 & 1807 & 1133 & 0.52 & 0.29 & 4.59 & 98.00\% \\
   \hline
$z_{(0)}$ & (varies) & 41.17 & 39.99 & 0.03 & 0.01 & 0.07 & 76.00\% \\
	\hline
	\hline
$\mathcal{R}_0$ & 3.50 & 3.92 & 3.76 & 0.18 & 0.06 & 0.95 & 95.00\% \\
   \hline
$\gamma$ & 0.30 & 0.31 & 0.29 & 0.21 & -0.01 & 1.16 & 99.00\% \\
   \hline
$S_{(0)}$ & 999 & 1909 & 1060 & 0.64 & 0.36 & 4.26 & 100.00\% \\
   \hline
$z_{(0)}$ & (varies) & 30.65 & 30.35 & 0.04 & -6.35\mbox{\sc{e}-3} & 0.05 & 45.00\% \\
	\hline
	\hline
$\mathcal{R}_0$ & 5.00 & 6.13 & 5.53 & 0.20 & 0.12 & 1.39 & 100.00\% \\
   \hline
$\gamma$ & 0.30 & 0.28 & 0.27 & 0.29 & -0.09 & 1.18 & 100.00\% \\
   \hline
$S_{(0)}$ & 999 & 2144 & 1220 & 0.68 & 0.49 & 4.94 & 99.00\% \\
   \hline
$z_{(0)}$ & (varies) & 26.28 & 25.26 & 0.03 & -0.01 & 0.03 & 52.00\% \\
	\hline
	\hline
	\hline
$\eta$ & Truth & Mean & Median & Error & Bias & Relative & 95\% HPD \\ 
&  &  &  &  &  &  HPD width & accuracy \\ 
	\hline
	\hline
$\mathcal{R}_0$ & 5.00 & 7.20 & 6.37 & 0.27 & 0.23 & 2.53 & 100.00\% \\
   \hline
$\gamma$ & 0.30 & 0.26 & 0.22 & 0.26 & -0.19 & 1.41 & 100.00\% \\
   \hline
$S_{(0)}$ & 9999 & 17,339 & 10,518 & 0.31 & 0.12 & 4.91 & 100.00\% \\
   \hline
$z_{(0)}$ & (varies) & 28.20 & 26.93 & 0.03 & -0.01 & 1.05 & 36.00\% \\
	\hline
\end{tabular}
\end{center}
{}
\label{table:sim}
\end{suppTable}

\clearpage

\newpage

\bibliographystyle{genetics}
\bibliography{volzSIR}

\end{document}